\documentclass[hyper]{JHEP3} 

\usepackage{epsfig,multicol}

\newcommand\fverb{\setbox\pippobox=\hbox\bgroup\verb}
\newcommand\fverbdo{\egroup\medskip\noindent%
                        \fbox{\unhbox\pippobox}\ }
\newcommand\fverbit{\egroup\item[\fbox{\unhbox\pippobox}]}
\newbox\pippobox

\def\bea{\begin{eqnarray}}
\def\eea{\end{eqnarray}}
\def\bec{\begin{center}}
\def\ec{\end{center}}

\def\beq{\begin{equation}}
\def\eeq{\end{equation}}

\def\l{\left}
\def\r{\right}

\title{Phenomenology of \\
Mixed Modulus-Anomaly Mediation\\
in Fluxed String Compactifications and Brane Models
}

\author{ Kiwoon Choi, Kwang-Sik Jeong\\
         Department of Physics, Korea Advanced Institute of Science
            and Technology\\
         Daejeon 305-701, Korea\\
        E-mail: \email{kchoi@hep.kaist.ac.kr},
        \email{ksjeong@hep.kaist.ac.kr} }
\author{ Ken-ichi, Okumura\\
         Department of Physics, Kyushu University, Fukuoka
            812-8581, Japan\\
        E-mail: \email{okumura@higgs.phys.kyushu-u.ac.jp} }

\preprint{KAIST-TH 05/05
}       

\abstract{
In some string compactifications, for instance the recently proposed
KKLT set-up, light moduli are stabilized  by
nonperturbative effects at supersymmetric AdS vacuum which is
lifted to a dS vacuum by supersymmetry breaking uplifting potential.
In such models, soft supersymmetry breaking terms
are determined by a specific mixed modulus-anomaly mediation
in which the two mediations  typically give  comparable contributions
 to soft parameters.
Similar pattern of soft terms can arise also in brane models to stabilize
the radion by nonperturbative effects.
We examine some phenomenological consequences of this mixed modulus-anomaly
mediation, including the pattern of low energy sparticle spectrum
and the possibility of electroweak symmetry breaking.
It is noted that adding the anomaly-mediated contributions
at $M_{GUT}$ amounts to replacing the messenger scale of the modulus mediation
by a  mirage messenger scale $(m_{3/2}/M_{Pl})^{\alpha/2}M_{GUT}$
where $\alpha=m_{3/2}/[M_0\ln(M_{Pl}/m_{3/2})]$ for $M_0$ denoting the modulus-mediated
contribution to the gaugino mass at $M_{GUT}$. The minimal KKLT set-up predicts $\alpha=1$.
As a consequence, for $\alpha={\cal O}(1)$,
the model can lead to a highly distinctive
pattern of sparticle masses at TeV scale,
particularly when $\alpha= 2$.
}

\keywords{Supergravity Models, Supersymmetry Breaking, Supersymmetric Standard Model}

\begin{document}


\section{Introduction}
Low energy supersymmetry (SUSY) is
one of the prime candidates for physics beyond the standard model at TeV scale \cite{Nilles:1983ge}.
One of the central questions in supersymmetric models is to
understand the mechanism of SUSY breaking, in particular the origin
of soft SUSY breaking terms in the low energy effective lagrangian \cite{kane}.
Most phenomenological aspects of supersymmetric models are determined
by those soft terms which would be induced by the auxiliary components of
SUSY breaking messenger fields.
In string theory, the most plausible candidates for messenger fields
are the moduli fields (including dilaton) describing the continuous degeneracy of string vacua
at leading  approximation \cite{modulimediation}.
In addition to  string moduli, the 4-dimensional supergravity (SUGRA)
multiplet provides a model-independent source of SUSY breaking,
i.e. the anomaly-mediation \cite{Randall:1998uk}, which generically induces the soft masses
 $m_{soft}\sim m_{3/2}/8\pi^2$
where $m_{3/2}$ is the gravitino mass.
To identify the dominant source of soft terms, one needs to understand how those moduli are stabilized
at a nearly 4D Poincare invariant vacuum.
For instance, if some moduli $\phi_H$ are stabilized with a heavy mass $m_{\phi_H}\gg 8\pi^2 m_{3/2}$,
those moduli would have the auxiliary component $F_{\phi_H}
\sim m_{3/2}^2/m_{\phi_H}\ll  m_{3/2}/8\pi^2$
(in the unit with $M_{Pl}=1$), thus their contribution to soft terms
are negligible even compared to the anomaly-mediated ones.
On the other hand, light moduli $\phi_L$ with a mass $m_{\phi_L}\sim m_{3/2}$ can have
$F_{\phi_L}\sim m_{3/2}$ which might provide $m_{soft}\sim m_{3/2}$
dominating over the anomaly-mediation.
In addition to stabilizing all the relevant moduli,
one needs to make sure that the resulting vacuum energy density is tuned to be
nearly zero since the soft scalar masses can be affected by any additional source
of the vacuum energy density \cite{choi3}.

Recently KKLT has proposed an interesting set-up to stabilize the moduli
within the framework of type IIB string theory.
The KKLT set-up \cite{Kachru:2003aw}
involves three steps
to achieve a SUSY breaking Minkowski (or de Sitter) vacuum, while stabilizing all
(or most of) moduli.
The first step  is to introduce the NS and RR 3-form fluxes,
$H_3$ and $F_3$,
stabilizing the dilaton $S$ and all complex structure moduli $Z_\alpha$.
For some flux vacua,  $G_3=H_3-iSF_3$ can be aligned nearly in the direction of  a primitive $(2,1)$-form,
for which $S$ and $Z_\alpha$ get superheavy masses not far below $M_{Pl}$, while the gravitino
remains to be light.
In the second step, one introduces nonperturbative dynamics, e.g. gaugino condensation
\cite{gauginocondensation},
to stabilize the K\"ahler moduli $T_i$.
This step fixes $T_i$ at an $N=1$ supersymmetric AdS vacuum with $m_T\approx m_{3/2}\ln(M_{Pl}/m_{3/2})$
and the vacuum energy density $V_0\approx-3m_{3/2}^2M_{Pl}^2$.
The last step is to introduce an anti-D3 brane ($\bar{D3}$) providing a
positive uplifting potential  which would make the total vacuum energy density to be
positive but small as desired. $\bar{D3}$ induces also a SUSY breaking vacuum shift which eventually
generates the soft SUSY breaking terms of visible fields \cite{choi1,choi2}.

The structure of soft terms in KKLT flux compactification has been studied in \cite{choi1,choi2}.
It has been noted that such compactification typically leads to
$F^T/T\sim m_{3/2}/4\pi^2$ (or even smaller in some special case)
and  $F^{S,Z}\ll F^T$,
implying that the loop-induced anomaly mediation
\cite{Randall:1998uk}
generically provides an important
contribution to soft terms.
If the visible gauge fields originate from $D3$ branes, the resulting soft terms
are dominated by the anomaly mediation whose phenomenology
has been extensively studied before \cite{anomalypheno}.
However in KKLT set-up, it is difficult to stabilize the position moduli
of $D3$ branes. Also the pure anomaly mediation suffers from the negative slepton
mass-square problem. In view of these difficulties,
a more attractive possibility is that
the visible gauge fields originate from $D7$ branes wrapping a 4-cycle.
In such case, the soft terms are induced by a specific mixed modulus-anomaly
mediation in which the two mediations give comparable contributions \cite{choi1,choi2}.

In fact, the KKLT set-up can be considered a specific example of more general scenario
in which the light moduli are stabilized by nonperturbative
dynamics yielding an $N=1$ SUSY AdS vacuum, and
this SUSY AdS vacuum is lifted to a SUSY-breaking Minkowski (or de Sitter) vacuum by an
appropriate uplifting mechanism which is assumed to be sequestered from the visible sector.
Such scenario might be realized also in some class of brane models.
Indeed the radion stabilization in 5D orbifold SUGRA
based on the boundary and bulk gaugino condensations \cite{luty}
can provide another example of this scenario which
results in the mixed radion-anomaly mediation.

In this paper we wish to examine some phenomenological aspects of this mixed modulus-anomaly
mediation, including the pattern of low energy sparticle spectrums
and the possibility of electroweak symmetry breaking.
The discussion will be made within the framework of
4D effective SUGRA with a SUSY breaking uplifting potential
which is sequestered from the visible sector.
As we will see, depending upon the anomaly to modulus mediation ratio $\alpha
=m_{3/2}/[M_0\ln(M_{Pl}/m_{3/2})]$ at the GUT scale $M_{GUT}$, where $M_0=F^T/(T+T^*)$ denotes
the modulus-mediated contribution to the gaugino mass at $M_{GUT}$,
the model can lead to a highly distinctive pattern of
superparticle masses at low energy scales.
This is essentially due to that the low energy soft parameters
in a mixed modulus-anomaly mediation with messenger scale
$\Lambda$ are (approximately) same as those
of the pure modulus-mediation with
a {\it mirage messenger scale} $\sim (m_{3/2}/M_{Pl})^{\alpha/2}\Lambda$.
The minimal KKLT model predicts $\alpha=1$, so has a mirage messenger scale
close to the intermediate scale $\sqrt{m_{3/2}M_{Pl}}$,
while the string, compactification and gauge unification
scales are all close to $M_{Pl}$.
If $\alpha=2$,
a striking pattern of low energy superparticle spectrum emerges
since the mirage messenger scale is close to TeV: soft masses appear to be unified at TeV
although the gauge couplings are unified at  $10^{16}-10^{17}$ GeV!
Although no string theory realization is found yet,
$\alpha=2$
can be naturally obtained by an uplifting mechanism to yield
an uplifting potential $V_{\rm lift}\propto 1/(T+T^*)$ \cite{choi1,choi2}.
($\bar{D3}$ in the KKLT set-up gives $V_{\rm lift}\propto 1/(T+T^*)^2$.)
Alternatively, one might be able to obtain a somewhat wide range of $\alpha$
(including $\alpha=2$) by tuning the form of the non-perturbative superpotential \cite{choi2}.
In the next section, we discuss  some features of the soft terms in mixed modulus-anomaly mediation,
which is largely based on the results of \cite{choi1,choi2}.
In sec. 3, we examine the resulting low energy soft parameters
and present the results of our phenomenological analysis.
Sec. 4 is the conclusion.

\section{Soft terms in mixed modulus-anomaly mediation}


To make a motivation for our study, let us start with a brief
discussion of soft terms in KKLT flux compactification following \cite{choi1,choi2}.
In KKLT models on CY orientifold,
the dilaton $S$ and complex structure moduli $Z_\alpha$ generically get  superheavy masses
of the order of compactification scale by the 3-form NS and RR fluxes.
This step of stabilizing $S$ and $Z_\alpha$ is assumed to preserve
the (approximate) $N=1$ SUSY, so the gravitino remains to be light
with $m_{3/2}\ll m_{S,Z}$.
To fix the K\"ahler moduli $T_i$, one  introduces
a superpotential of the form $W_{np} = Ae^{-aT_i}$ induced by gaugino
condensations on $D7$ branes.
Since $m_{Z,S}\gg m_{3/2}$,
the stabilization of $T_i$ and also the low energy SUSY breaking
can be described by an effective SUGRA obtained after integrating out
 $S$ and $Z_\alpha$.
For simplicity, here we consider only the case with single  K\"ahler modulus $T$
as the generalization to multi K\"ahler moduli is rather straightforward.
Then the effective SUGRA of $T$ and the gauge and matter superfields on $D7/D3$
can be written as
\begin{eqnarray}
\label{N=1}
S_{N=1}&=&\int d^4x \sqrt{g^C} \,\left[\,
\int d^4\theta \,
CC^*\left(-3\exp(-K_{eff}/3)\right)
\right.\nonumber \\
&&+\,\left.\left\{
\int d^2\theta
\left(\frac{1}{4}f_a W^{a\alpha}W^a_\alpha
+C^3W_{eff}\right)
+{\rm h.c.}\right\}\,\right],
\end{eqnarray}
where
\bea
K_{eff}&=&K_0(T+T^*)+Z_i(T+T^*)Q_i^*Q_i,
\nonumber \\
W_{eff}&=& W_0(T)+\frac{1}{6}\lambda_{ijk}Q_iQ_jQ_k.
\eea
Here $g^C_{\mu\nu}$ is the 4D metric
in superconformal frame which is related to the Einstein frame metric
$g^E_{\mu\nu}$ as
$g^C_{\mu\nu}=(CC^*)^{-1}e^{K_{eff}/3}g^E_{\mu\nu}$,
$C=C_0+F^C\theta^2$ is the chiral compensator superfield
of 4D $N=1$ SUGRA, and $Q_i$ are the gauge-charged matter superfields.

The modulus K\"ahler and superpotential of the minimal KKLT set-up are given by
\bea
\label{minimal}
K_0&=&-3\ln(T+T^*),
\nonumber \\
W_0&=& w_0-Ae^{-aT},
\eea
where the constant piece $w_0$ of the superpotential originates from
the fluxes.
Using the following $U(1)_R$ transformation of the superconformal formulation of
the 4D $N=1$ SUGRA:
\bea
\label{u1r}
U(1)_R: \quad
C\rightarrow e^{i\beta_R}C,
\quad W_{eff}\rightarrow e^{-3i\beta_R} W_{eff},
\eea
one can make $w_0$ to be a real positive
constant without loss of generality.
The K\"ahler potential (and also the uplifting potential which will be introduced
later) of the KKLT model possesses an approximate nonlinear PQ  symmetry,
\bea
\label{nonlinearpq}
U(1)_T: \quad
T\rightarrow T+i\beta_T,
\eea
with which one can make $A$ to be a real positive constant
again without loss of generality.
As was noticed before \cite{choi2,choi5}, this nonlinear PQ symmetry is crucial
for the KKLT set-up to avoid dangerous SUSY CP violation.

The holomorphic Yukawa couplings $\lambda_{ijk}$ are independent of $T$,
however the matter K\"ahler metric and holomorhic gauge kinetic functions can have
nontrivial $T$-dependence as
\bea
\label{kahlermetric}
Z_i&=& \frac{1}{(T+T^*)^{n_i}},
\nonumber \\
f_a&=&T^{\, l_a},
\eea
where $n_i=0$ and $l_a=1$ for the matter and gauge fields living on $D7$,
while $n_i=1$ and $l_a=0$ for the matter and gauge fields on $D3$
\cite{ibanez1}.
When the matter fields live on the intersections of $D7$ branes, $n_i$ can have a fractional
value, e.g. $n_i=1/2$ \cite{ibanez1,ibanez}.

The modulus superpotential $W_0$
stabilizes $T$ at
\begin{eqnarray}
\label{tstabilization}
\langle  aT\rangle \,\approx\,\ln(A/w_0)\,\approx\, \ln (M_{Pl}/m_{3/2}),
\end{eqnarray}
however the resulting ground state is a SUSY preserving AdS vacuum.
To obtain a SUSY-breaking Minkowski (or dS) vacuum, KKLT proposed
to add an anti-${D3}$ brane ($\bar{D3}$)  providing a positive uplifting potential.
Such $\bar{D3}$ is stabilized at the tip of a smoothed conifold singularity
at which the geometry is highly warped with
an exponentially small warp factor $e^{A_{min}}\sim \sqrt{m_{3/2}/M_{PL}}$
\cite{giddings}.
On $\bar{D3}$,  $N=1$ SUSY is broken  explicitly or is non-linearly realized \cite{nonlinear}.
It has been argued \cite{choi2} that the low energy consequence of ${\bar{D3}}$ in KKLT set-up
can be described by a single spurion operator
up to small corrections further suppressed by $e^{A_{min}}$:
\begin{eqnarray}
\label{N=0}
S_{\rm lift}&=&-\int d^4x \sqrt{g^C} \int d^4\theta \,
\,C^2C^{2*}\theta^2
\bar{\theta}^2\,\,{\cal P}_{\rm lift}(T,T^*),
\end{eqnarray}
where
\bea
\label{lifting}
{\cal P}_{\rm lift}=D(T+T^*)^{n_P}
\eea
for a
positive constant $D={\cal O}(e^{4A_{min}}M_{Pl}^4)=
{\cal O}(m_{3/2}^2M_{Pl}^2)$.
Including this spurion operator, the low energy effective action of
KKLT compactification
is given by
\bea
\label{effectiveaction}
S_{eff}=S_{N=1}+S_{\rm lift}.
\eea
From this, one finds
that the SUSY breaking order parameters (in the Einstein frame) approximately
take the standard $N=1$ SUGRA form:
\begin{eqnarray}
\label{approx-F}
\frac{F^C}{C_0}&= &\frac{1}{3}\partial_TK_0F^T+e^{K_0/2}W_0^*,
\nonumber \\
F^T&= &-e^{K_0/2}K_0^{TT^*}\left(D_T
W_0\right)^*,
\nonumber \\
m_{3/2}&= & e^{K_0/2}W_0,
\end{eqnarray}
while the modulus potential contains the uplifting term
\begin{eqnarray}
\label{approx-potential}
V_0&=&e^{K_0}\left(K^{TT^*}_0D_TW_0(D_TW_0)^*-3|W_0|^2\right)+V_{\rm lift},
\end{eqnarray}
where
\bea
\label{uplift}
V_{\rm lift}=e^{2K_0/3}{\cal P}_{\rm lift}(T,T^*) \equiv \frac{D}{(T+T^*)^{2-n_P}}.
\eea
It is now straightforward to compute the SUSY breaking order parameters
$F^T$ and $F^C$ (or equivalently $m_{3/2}$)  by
minimizing the above modulus potential under the fine tuning
for $\langle V_0\rangle =0$.
One finds \cite{choi1,choi2}
\begin{eqnarray}
\label{C/T}
\frac{F^C}{C_0}\,\approx\, m_{3/2}&\approx& \frac{w_0}{M_{Pl}^2(T+T^*)^{3/2}},
\nonumber \\
\frac{F^T}{(T+T^*)}&\approx& \frac{2-n_P}{a(T+T^*)}m_{3/2}.
\end{eqnarray}
For the uplifting potential originating from $\bar{D3}$,
the corresponding ${\cal P}$ is a $T$-independent constant, i.e. $n_P=0$
\cite{burgess,kkmmlt}.

One of the most interesting features of the KKLT flux compactification
is that
\bea
\label{mixed}
M_0\,\equiv\,\frac{F^T}{(T+T^*)}={\cal O}\left(\frac{m_{3/2}}{4\pi^2}\right)
\eea
for $m_{3/2}$ near the TeV scale,
which suggests that the loop-induced anomaly mediation and
the tree-level modulus mediation can be comparable to each other.
This is essentially due to the relation
$aT \approx \ln(M_{Pl}/m_{3/2})={\cal O}(4\pi^2)$.
As we will see, phenomenological consequences of the mixed modulus-anomaly mediation
are somewhat sensitive to the ratio between the anomaly and modulus mediations
which we will parameterize by
\bea
\label{alpha}
\alpha &\equiv& \frac{m_{3/2}}{M_0\ln(M_{Pl}/m_{3/2})}\,\approx\,
\frac{2}{a(T+T^*)}\frac{F^C}{C_0}
\frac{(T+T^*)}{F^T}.
\eea
The minimal  KKLT set-up described by the modulus K\"ahler and superpotential
(\ref{minimal}) and the uplifting potential (\ref{uplift}) with $n_P=0$
predicts
\bea
\left.\alpha\right|_{_{KKLT}}=1+{\cal O}\left(\frac{1}{\ln(M_{Pl}/m_{3/2})}\right).
\eea
However one might be able to generalize
the model to obtain a different value of $\alpha$.
It has been noticed \cite{choi2} that a model with
racetrack superpotential
$W_0=-A_1e^{-a_1T}+A_2e^{-a_2T}$ \cite{racetrack} can give a stable Minkowski vacuum
with a light gravitino and $\alpha={\cal O}(4\pi^2)$ when $(a_2-a_1)/(a_2+a_1)={\cal O}(1/4\pi^2)$.
Motivated by this observation,
one can consider a more general class of effective SUGRA described by
\bea
\label{generalmodel}
K_0&=&-n_0\ln(T+T^*),
\nonumber \\
W_0&=&w_0-A_1e^{-a_1T}+A_2e^{-a_2T} \quad (a_1\leq a_2),
\nonumber \\
{\cal P}_{\rm lift}&=&D\,(T+T^*)^{n_P},
\eea
for which
\bea
V_{\rm lift}=e^{2K_0/3}{\cal P}_{\rm lift}=\frac{D}{(T+T^*)^{\frac{2}{3}n_0-n_P}}.
\eea
For the parameter regions which give a stable Minkowski vacuum with light gravitino,
we have examined the value of $\alpha$ predicted by this model, and found
\bea
\label{alpha1}
\alpha\,=\,\frac{\xi}{1-3n_P/2n_0}\,,
\eea
where $\xi$ is close to 1 in most of the parameter spaces, but
can be significantly bigger than 1 for
$|w_0/A_1e^{-a_1T}|\ll 1$ and $(a_2-a_1)/(a_2+a_1)={\cal O}(1/4\pi^2)$
as anticipated in \cite{choi2}.
If the uplifting spurion operator
${\cal P}_{\rm lift}$ is a $T$-independent constant as in the KKLT case,
i.e. $n_P=0$, we have $\alpha=\xi$.
Although no concrete realization is found yet,
string theory might be able to provide other forms  of ${\cal P}_{\rm lift}$,
e.g. $n_P=1$ or $n_P=-1$ which gives
$\alpha=2\xi$ or $\alpha=2\xi/3$ for $n_0=3$.
In this paper, we simply take this possibility without questioning the origin of
${\cal P}_{\rm lift}$, and treat $\alpha$ as a free parameter
while focusing on $\alpha={\cal O}(1)$ for which the resulting phenomenology is
most interesting.
Note that $\alpha\ll 1$  corresponds to the limit of pure modulus-mediation,
while $\alpha\gg 1$ corresponds to the pure anomaly mediation.

Let us now consider the soft SUSY breaking terms of  canonically normalized
visible fields for generic value of $\alpha$:
\begin{eqnarray}
{\cal L}_{soft}&=&-\frac{1}{2}M_a\lambda^a\lambda^a-m_i^2|\tilde{Q}_i|^2
-\frac{1}{6}A_{ijk}y_{ijk}\tilde{Q}_i\tilde{Q}_j\tilde{Q}_k+{\rm h.c.},
\end{eqnarray}
where $\lambda^a$ are gauginos, $\tilde{Q}_i$ are sfermions,
and $y_{ijk}$ denote the canonically normalized Yukawa couplings:
\begin{eqnarray}
y_{ijk}=\frac{\lambda_{ijk}}{\sqrt{e^{-K_0}Z_iZ_jZ_k}}.
\end{eqnarray}
For $\alpha={\cal O}(1)$,
the loop-induced anomaly-mediation \cite{Randall:1998uk}
becomes comparable  to the modulus mediation, thus should be included in
the soft terms at scales below the compactification
(unification) scale.
This results in a  mixed modulus-anomaly mediation, yielding the following form of
soft parameters at energy scale just below the unification scale \cite{choi1,choi2}:
\begin{eqnarray}
\label{soft1}
M_a&=& F^T\partial_T\ln\left({\rm Re}(f_a)\right) +\frac{b_ag_a^2}{8\pi^2}\frac{F^C}{C_0}
\nonumber \\
&=& l_aM_0+\frac{b_a}{8\pi^2}g^2_{GUT}m_{3/2},
\nonumber \\
A_{ijk}&=&
-F^T\partial_T\ln\left(\frac{\lambda_{ijk}}{e^{-K_0}Z_iZ_jZ_k}\right)-
\frac{1}{16\pi^2}(\gamma_i+\gamma_j+\gamma_k)\frac{F^C}{C_0}
\nonumber \\
&=&a_{ijk}M_0
-\frac{1}{16\pi^2}(\gamma_i+\gamma_j+\gamma_k)m_{3/2},
\nonumber \\
m_i^2&=&
\frac{2}{3}V_0
-F^TF^{T*}\partial_T\partial_{T^*}
\ln \left(e^{-K_0/3}Z_i\right)
-\frac{1}{32\pi^2}\frac{d\gamma_i}{d\ln\mu}\left|\frac{F^C}{C_0}\right|^2
 \nonumber \\
&&
+ \frac{1}{16\pi^2}\left\{ (\partial_{T}{\gamma}_i)
F^T\left(\frac{F^C}{C_0}\right)^*
+{\rm h.c.}\right\}
\nonumber \\
&=&c_i|M_0|^2
-\frac{1}{32\pi^2}\frac{d\gamma_i}{d\ln\mu}|m_{3/2}|^2
\nonumber\\
&&
+\frac{1}{8\pi^2}\left\{
\sum_{jk}a_{ijk}\left|\frac{y_{ijk}}{2}\right|^2-
\sum_A l_Ag_A^2C_A(Q_i)\right\}
\left(M_0m_{3/2}^*
+M_0^*m_{3/2}\right),
\end{eqnarray}
where
\bea
a_{ijk}=3-n_i-n_j-n_k,
\quad
c_i=1-n_i
\nonumber
\eea
for the matter K\"ahler metric
$Z_i=1/(T+T^*)^{n_i}$,
\bea
C_A(Q_i){\bf 1}= \sum_{a\in A} T_a^2(Q_i)
\nonumber
\eea
for the $A$-th gauge group,
and
$M_0=F^T/(T+T^*)$.
Here $b_a$ and $\gamma_i$ are the one-loop
beta function coefficients and the anomalous dimension of $Q_i$, respectively,
defined by $\frac{dg_a}{d\ln \mu}=\frac{b_a}{8\pi^2} g_a^3$ and
$\frac{d\ln Z_i}{d\ln \mu}=\frac{1}{8\pi^2}\gamma_i$:
\bea
b_a&=&-\frac{3}{2}{\rm tr}\left(T_a^2({\rm Adj})\right)+\frac{1}{2}\sum_i
{\rm tr}\left(T^2_a(Q_i)\right),
\nonumber \\
\gamma_i&=&2\sum_Ag_A^2C_A(Q_i)-\frac{1}{2}\sum_{jk}|y_{ijk}|^2,
\nonumber \\
\partial_{T}\gamma_i
&=&-\frac{1}{2}\sum_{jk}|y_{ijk}|^2\partial_T\ln\left(\frac{\lambda_{ijk}}{e^{-K_0}Z_iZ_jZ_k}\right)
-2\sum_A g_A^2C_A(Q_i)\partial_T\ln\left({\rm Re}(f_A)\right).
\nonumber
\eea
Here we have ignored the off-diagonal terms of
$\omega_{ij}=\sum_{kl}y_{ikl}y^*_{jkl}$.

If the visible gauge fields originate from $D3$ branes, so $l_a=0$ and $n_i=1$,
the resulting soft terms correspond to the pure anomaly mediation whose phenomenology
has been extensively studied before \cite{anomalypheno}.
However in KKLT set-up, it is difficult to stabilize the position moduli
of those $D3$ branes. Also the pure anomaly mediation suffers from the negative slepton
mass-square problem. In view of these difficulties,
a more  attractive possibility is that
the visible gauge fields originate from $D7$ branes,
for which $l_a=1$ but still $n_i$ can be either 0 or 1/2 or even 1,
depending on the  origin of $Q_i$.
In the following, we will set $l_a=1$, and then
the generic mixed modulus-anomaly mediation is parameterized by
\bea
M_0, \,\, a_{ijk}, \,\, c_i, \,\, \alpha=m_{3/2}/[M_0\ln(M_{Pl}/m_{3/2})],
\nonumber
\eea
where the first three parameters are determined by
the modulus-dependence of the matter K\"ahler metric, while
$\alpha$ is determined by the mechanism of modulus stabilization
and the subsequent uplifting.

For the minimal KKLT set-up defined by (\ref{minimal}), (\ref{kahlermetric})
and (\ref{lifting}), the invariance of
the matter K\"ahler metric (\ref{kahlermetric})
under the nonlinear PQ symmetry (\ref{nonlinearpq}) assures
that $a_{ijk}$ are real.
We already noticed that the $U(1)_R$ transformation (\ref{u1r})
and the nonlinear PQ transformation (\ref{nonlinearpq}) can be used
to make the two parameters $w_0$ and $A$
in the modulus superpotential (\ref{minimal}) to be real without loss
of generality.
In such field basis, the resulting $F^T$ and $F^C$ are real,
thus the gaugino masses and $A$-parameters
in the minimal KKLT set-up do not contain any dangerous
CP-violating phase \cite{choi2,choi5}.

For a later discussion of the electroweak symmetry breaking,
let us  discuss the Higgs mass parameters
$\mu$ and $B$ in the mixed modulus-anomaly mediation scenario.
(Here we are using the same notation $\mu$ for both the Higgsino mass parameter
and the renomalization point.)
One possible source of  $\mu$ and $B$  is the Higgs bilinear terms
in K\"ahler and superpotential, which would take the following form
\cite{giudice,chun}
\bea
\label{higgsbilinear}
\Delta K_{eff}&=& \frac{H_1H_1^*}{(T+T^*)^{n_{H_1}}}
+\frac{H_2H_2^*}{(T+T^*)^{n_{H_2}}}+
\left(\frac{\kappa H_1H_2}{(T+T^*)^h} +{\rm h.c.}\right),
\nonumber \\
\Delta W_{eff}&=& \tilde{A}e^{-aT}H_1H_2,
\eea
where $\Delta W_{eff}$ is induced by the non-perturbative
dynamics yielding $Ae^{-aT}$ in the modulus superpotential
(\ref{minimal}).
The resulting $\mu$ for the canonically normalized Higgs doublets
at the unification scale
is given by
\bea
\label{muandb}
\mu = \mu_W+\mu_K\,,
\eea
where
\bea
\mu_W=\frac{\tilde{A}e^{-aT}}{(T+T^*)^{l_W}},
\quad
\mu_K=\frac{\kappa}{(T+T^*)^{l_K}}
\left(\frac{F^C}{C_0}+(1-h)\frac{F^T}{(T+T^*)}\right)^*
\eea
for
$l_W=(3-n_{H_1}-n_{H_2})/2$ and $l_K=(2h-n_{H_1}-n_{H_2})/2$.
Soft masses in the mixed modulus-anomaly mediation are of the order of $M_0=F^T/(T+T^*)$,
thus it is desirable that $\mu$ is ${\cal O}(M_0)$ also.
Although $F^C/C_0\approx m_{3/2}={\cal O}(4\pi^2 M_0)$ in the mixed modulus-anomaly mediation,
it is not difficult to obtain $\mu={\cal O}(M_0)$.
In view of that $\tilde{A}\approx A/M_{Pl}^2$ and
 $Ae^{-aT}\approx m_{3/2}M_{Pl}^2/aT$ for
the modulus superpotential (\ref{minimal}),
$\mu_W$ is
naturally of the order of $M_0$.
As for $\mu_K$, one might assume that the $H_1H_2$-term
in $K_{eff}$ is induced by a loop correction, and thus
$\mu_K={\cal O}(m_{3/2}/4\pi^2)={\cal O}(M_0)$.

In fact, the real problem is to get the desired size of $B$.
For $B$ originating from (\ref{higgsbilinear}), we find
\bea
B\mu &=&- \left[m_{3/2}-a(T+T^*)M_0+{\cal O}(M_0)\right]\mu_W
+\left[m_{3/2}+{\cal O}(M_0)\right]\mu_K
\nonumber \\
&=& m_{3/2}\big[\,\mu_K+\left(\frac{2}{\alpha}-1\right)\mu_W\,\big]+{\cal O}(M_0^2)
\eea
where we have used $\alpha\approx 2m_{3/2}/a(T+T^*)M_0$.
This result shows that $B$ from (\ref{higgsbilinear}) is generically of the order of $m_{3/2}$, thus
would be too large to achieve the correct electroweak symmetry breaking.
Unless $\alpha\approx 1$, one can still obtain the desired $B={\cal O}(M_0)$ under
the fine-tuning:
\bea
\label{b-tuning}
\mu_K+\left(\frac{2}{\alpha}-1\right)\mu_W={\cal O}\left(\frac{\mu_K+\mu_W}{4\pi^2}\right).
\eea
which we will assume in the discussion of electroweak symmetry breaking
in the next section.
We note that for the models  yielding $\alpha\approx 2$,
for instance a model with $n_0=3, n_P=1, A_2=0$ in (\ref{generalmodel})
which gives $\alpha=2+{\cal O}(1/4\pi^2)$,
the above condition is automatically satisfied
when $\mu_K=0$, i.e. when
the $\mu$ term originates entirely from the nonperturbative term in $W_{eff}$.
For the case with $\alpha\approx 1$, which would be the most interesting case as
it is predicted by the minimal KKLT set-up, the
fine-tuning (\ref{b-tuning}) is not allowed.
One might then consider
\bea
\label{kkltmu}
\Delta K_{eff}&=& \frac{H_1H_1^*}{(T+T^*)^{n_{H_1}}}
+\frac{H_2H_2^*}{(T+T^*)^{n_{H_2}}}+
\left(\frac{\kappa H_1H_2}{(T+T^*)^h} +{\rm h.c.}\right),
\nonumber \\
\Delta W_{eff}&=& \tilde{\mu}H_1H_2,
\eea
where $\tilde{\mu}$ is a $T$-independent
constant which is adjusted to give $\mu_W=\tilde{\mu}/(T+T^*)^{l_W}={\cal O}(M_0)$.
In this case, one easily finds
$\mu=\mu_K+\mu_W$ and $B\mu=m_{3/2}(\mu_K-\mu_W)+{\cal O}(M_0^2)$,
thus $B={\cal O}(M_0)$ can be obtained
through the fine-tuning:  $\mu_K-\mu_W={\cal O}(\mu/4\pi^2)$.

Generating the $\mu$ and $B$-terms
through (\ref{higgsbilinear}) or (\ref{kkltmu}) has another unattractive feature
in addition to the involved fine-tuning to get $B={\cal O}(M_0)$.
Those models for $\mu$ and $B$ involve
a CP-violating phase ${\rm Arg}(\tilde{A}\kappa^*)$ or ${\rm Arg}(\tilde{\mu}\kappa^*)$
which eventually generates a nonzero ${\rm Arg}(B)$. To satisfy the constraints
from the hadron or electron electric dipole moments, one then needs
to tune this CP phase to be smaller than $10^{-2}$.

The difficulty to obtain $B={\cal O}(M_0)$
is a generic problem of models which predict $m_{3/2}\gg M_0$.
A simple way to avoid this difficulty is to
assume that the Higgs $\mu$-term originates from a trilinear Yukawa term
involving a singlet $N$ \cite{Ellis:1988er}:
\bea
\label{nmssm}
\Delta W_{eff}= \lambda_1 NH_1H_2+\frac{\lambda_2}{3} N^3.
\eea
In this case, the (effective) $\mu$ and $B$ are given by
\bea
\mu&=&e^{K_0/2}\lambda_1 \langle N \rangle,
\nonumber \\
B & = & A_{NH_1H_2}+e^{K_0/2}\lambda_2^* \langle N \rangle/Z_N,
\eea
where $Z_N$ is the K\"ahler metric of $N$, thus they have the
desired size of ${\cal O}(M_0)$.
In addition to giving the correct size of $\mu$ and $B$ without a
fine-tuning of parameters, this model has another virtue:
it avoids naturally all the potentially dangerous CP phases in the soft
parameters.
Using the field redefinitions
\bea
U(1)_H : \, H_1H_2 \rightarrow e^{i\beta_H}H_1H_2,
\quad
U(1)_N : \, N \rightarrow e^{i\beta_N} N,
\eea
one can make $\lambda_1$ and $\lambda_2$ to be real, for which
$\langle N\rangle$ is real also.
We already noticed that the field redefinitions of
(\ref{u1r}) and (\ref{nonlinearpq}) can be used to assure
that the gaugino masses and $A$-parameters in the minimal KKLT set-up
are all real. When the Higgs $\mu$ and $B$ parameters originate
from the superpotential (\ref{nmssm}), the resulting $B$ is
automatically real, thus the model is completely free from
the potentially dangerous SUSY CP violation.

Before closing this section, we note that the mixed modulus-anomaly mediation can arise
also from 5D brane models on $S^1/Z_2$ stabilizing the radion by nonperturbative
effects \cite{luty}.
To see this, let us first note that gauge fields propagating in 5D bulk have
$f_a=T$ for the radion superfield $T=R+iB_5$
where $R$ is the orbifold radius and $B_5$ is the fifth component of the 5D graviphoton
\cite{marti}.
On the other hand, for gauge fields confined in the boundaries
of $S^1/Z_2$, the corresponding $f_a$ are $T$-independent constants.
The radion K\"ahler potential is given by $K_0=-3\ln(T+T^*)$,
and a superpotential of the form $W_0=w_0-\sum_iA_ie^{-a_iT}$
can be generated by gaugino condensations. The constant piece $w_0$ would be induced
by a gaugino condensation at the boundary, while $\sum_iA_ie^{-a_iT}$
arises from bulk gaugino condensations.
In the simplest situation, the matter K\"ahler metric takes the form of (\ref{kahlermetric}),
$n_i=1$ for boundary matter fields and $n_i=0$ for bulk matter fields \cite{marti}.
In fact, if a bulk matter field $Q_i$ have nonzero  5D mass $M_i$,
its zero mode has a K\"ahler metric given by $Z_i=(1-e^{M_i(T+T^*)})/M_i(T+T^*)$
\cite{choi4}.
In such case, the resulting Yukawa couplings
$y_{ijk}=\lambda_{ijk}/\sqrt{e^{-K_0}Z_iZ_jZ_k}$ can have a hierarchical structure
in a natural manner.
Again the radion superpotential $W_0$ from gaugino condensations can stabilize $T$
at a SUSY AdS vacuum.
Adding a SUSY-breaking anti-brane at the boundary will uplift
this AdS vacuum to a Minkowski (or dS) vacuum.
The corresponding spurion operator  ${\cal P}_{\rm lift}$ is a $T$-independent constant, so $n_P=0$
\cite{luty}.
If ${\cal P}_{\rm lift}$ originates dominantly from a
$T$-dependent bulk $U(1)$ FI-term
$D_{FI}\propto \partial_TK_0$ \cite{burgess},
one would have
${\cal P}_{\rm lift}=e^{-2K_0/2}V_D\propto g^2e^{-2K_0/3}(\partial_TK_0)^2 \propto 1/(T+T^*)$
so $\alpha\approx 2/3$.
On the other hand, if ${\cal P}_{\rm lift}$ originates dominantly from a
$T$-independent $U(1)_R$ FI term \cite{abe}, ${\cal P}_{\rm lift}\propto g^2e^{-2K_0/3} \propto
(T+T^*)$ and then $\alpha\approx 2$.
However, as was pointed out in \cite{choi2},  $U(1)$ FI-dominated
uplifting is difficult to be achieved since the FI D-term always vanishes
for a SUSY AdS solution of the $F$-term potential.


\section{Low energy phenomenology}

In this section, we examine some phenomenological consequences of the mixed
modulus-anomaly mediation given by (\ref{soft1}) with
$l_a=1$ and $\alpha$ in the range of ${\cal O}(1)$.
For concrete analysis, we will use the standard value of the unification scale
$M_{GUT}\sim 2\times 10^{16}$ GeV.

Mixed modulus-anomaly mediation can give a low energy sparticle
spectrum which is quite different from other scenarios of SUSY breaking.
This is mainly due to the particular correlation between the anomaly mediation
and the RG evolution of soft parameters.
To see this, let us consider the low energy gaugino masses.
At a scale $M^{(-)}_{GUT}$ just below $M_{GUT}$, the gaugino masses are given by
\bea
M_a(M^{(-)}_{GUT})=M_0\left(1+\frac{\ln(M_{Pl}/m_{3/2})}{8\pi^2}\,\alpha b_ag_{GUT}^2\right),
\eea
where $\alpha$ is defined in (\ref{alpha}).
The one-loop RG evolution between $M^{(-)}_{GUT}$ and $\mu$
yields
\bea
\label{lowgaugino}
M_a(\mu)=M_0\left[
1-\frac{1}{4\pi^2}b_ag_a^2(\mu)\ln\left(\frac{M_{GUT}}{(M_{Pl}/m_{3/2})^{\alpha/2}\mu}\right)
\right].
\eea
This result shows an interesting feature:
if $\alpha={\cal O}(1)$ as in KKLT set-up,
the anomaly-mediated contribution, i.e. the $\alpha$-dependent part, cancels significantly
the RG evolution of the modulus-mediated gaugino masses.
In particular it shows that the low energy gaugino masses in the
mixed modulus-anomaly mediation started from the messenger scale
$M_{GUT}$ are same as
the low energy gaugino masses in the pure modulus-mediation
started from a {\it mirage messenger scale} $\sim (m_{3/2}/M_{Pl})^{\alpha/2}M_{GUT}$.
Note that this mirage messenger scale does not correspond to a physical threshold
scale. Still the physical gauge coupling unification scale
is $M_{GUT}$, and the Kaluza-Klein and string threshold scales
are a little above $M_{GUT}$.
When $\alpha \approx 2$,
$M_a(M_{SUSY})$ are approximately same as
the pure modulus-mediated gaugino mass without any RG running effect,
\bea
M_a(M_{SUSY})\approx M_0\quad
\mbox{for}\quad
\alpha\approx 2.
\eea
In this case, we have unified  gaugino masses
at TeV, while the corresponding gauge couplings  are unified at $M_{GUT}\sim
2\times 10^{16}$ GeV.
%
\begin{figure}[t]
\begin{center}
\begin{minipage}{15cm}
\centerline{
{\hspace*{-.2cm}\psfig{figure=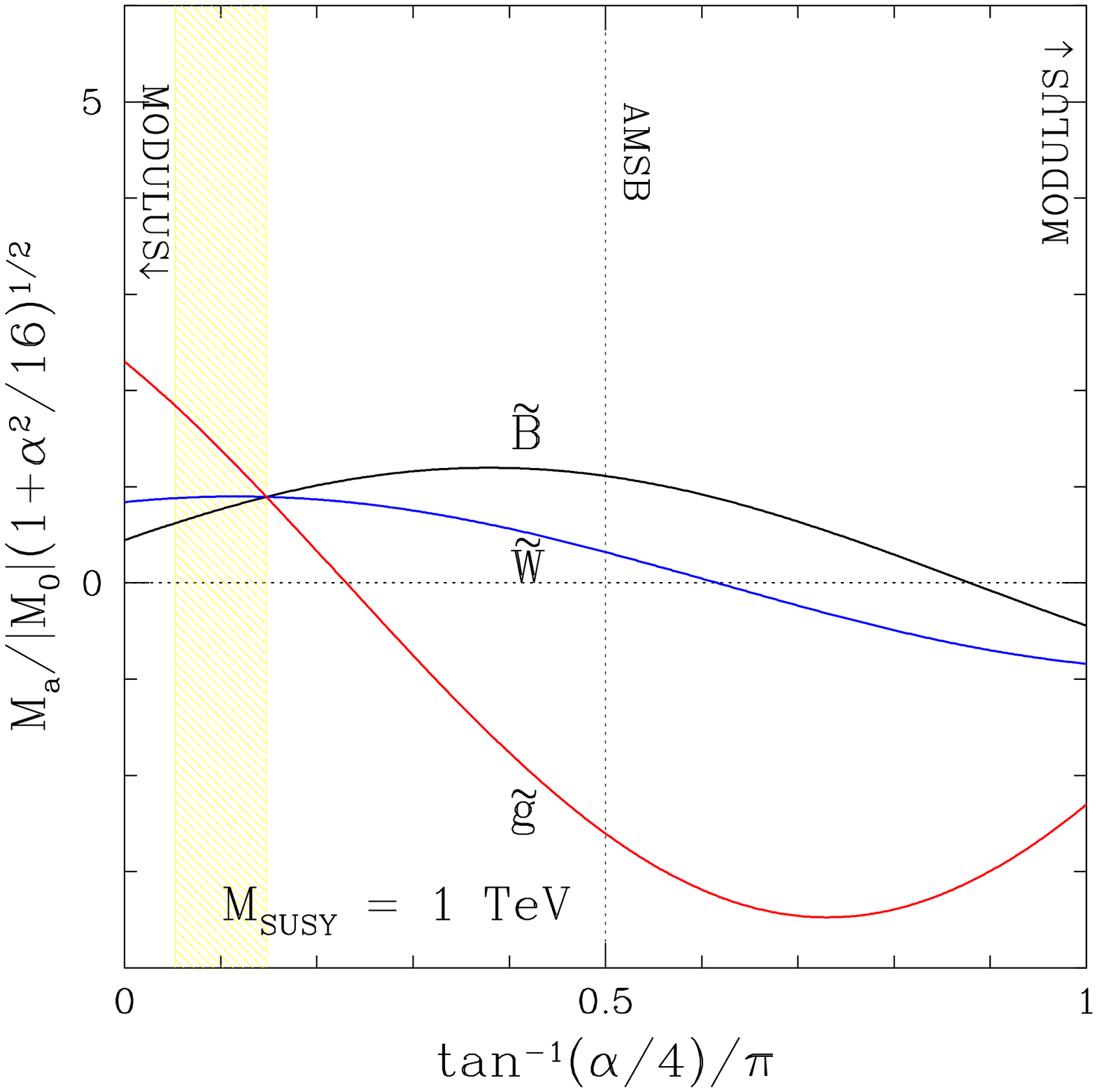,angle=0,width=7.5cm}}
{\hspace*{-.2cm}\psfig{figure=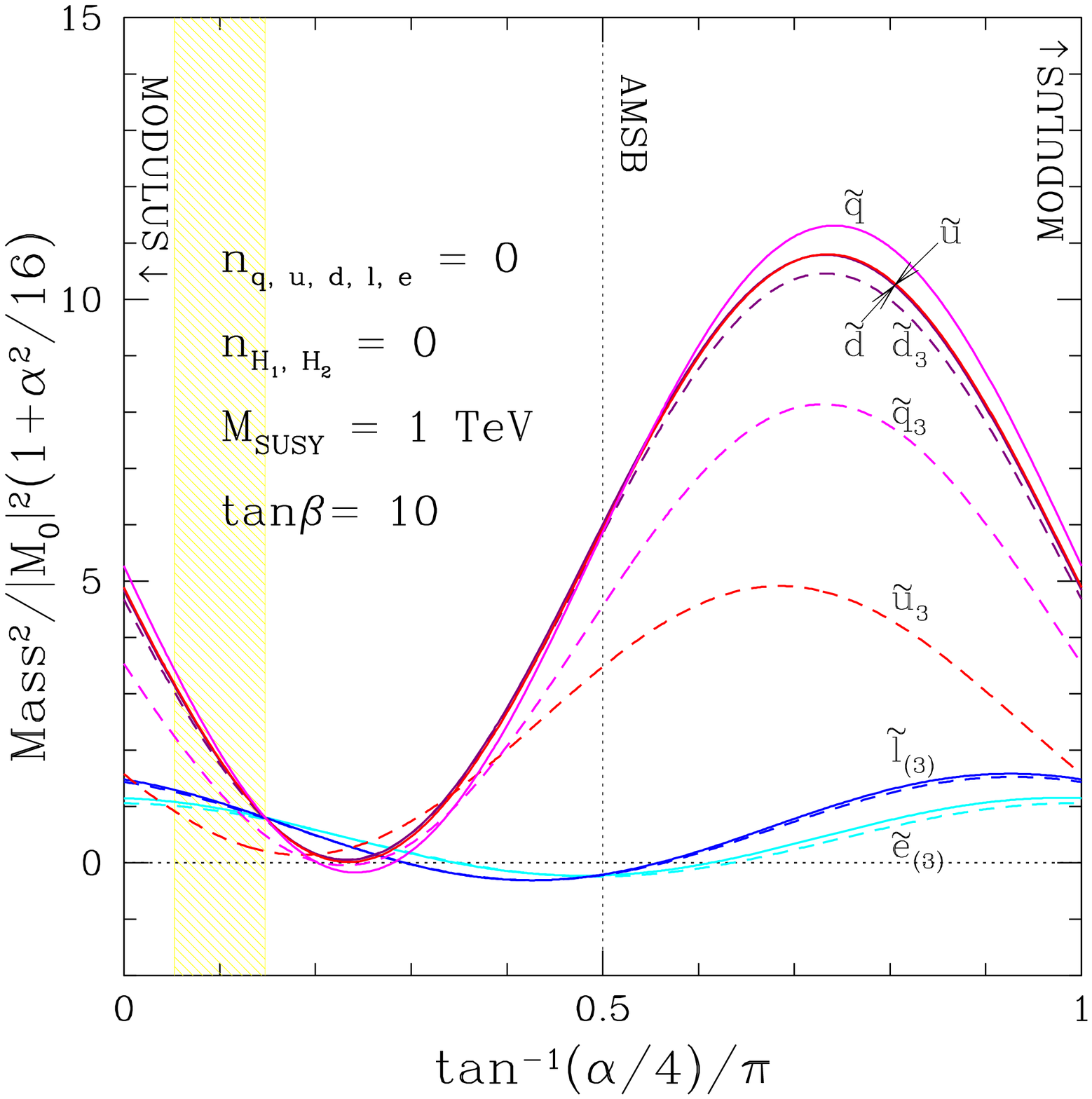,angle=0,width=7.5cm}}
}
\caption{Sparticle mass spectrum (relative ratios) at $M_{SUSY}=1$ TeV
for the entire range of the anomaly to modulus ratio $\alpha=m_{3/2}/[M_0\ln(M_{Pl}/m_{3/2})]$.
Here $M_0=F^T/(T+T^*)$ and $m_{3/2} =F^C/C_0$.
The shaded region indicates the range $2/3\leq \alpha\leq 2$
and the short-dashed curves denote the 3rd generation squarks/sleptons.
Note that the sign convention of the gaugino masses (and $A_{ijk}$) for $0\leq \tan(\alpha/4)\leq
\pi/2$ is different from the convention for
$\pi/2\leq \tan(\alpha/4)\leq \pi$.
\label{fig:tree_rewsb}}
\end{minipage}
\end{center}
\end{figure}
%
The low energy values of $A_{ijk}$ and $m_i^2$
show a similar feature.
In fact, if $y_{ijk}$ are non-vanishing {\it only}
for the combinations $Q_iQ_jQ_k$ satisfying $a_{ijk}=1$ and $c_i+c_j+c_k=1$,
{\it or} if the effects of the Yukawa couplings on the renormalization group
evolution can be ignored,
the resulting $A_{ijk}$ and $m_i^2$ at low energies are given by
\bea
\label{lowsfermion}
A_{ijk}(\mu)&= &M_0\left[a_{ijk}
+\frac{1}{8\pi^2}(\gamma_i(\mu)
+\gamma_j(\mu)+\gamma_k(\mu))
\ln\left(\frac{M_{GUT}}{(M_{Pl}/m_{3/2})^{\alpha/2}\mu}\right)\right],
\nonumber \\
m_i^2(\mu)&= & \left|M_0\right|^2\left[
c_i+\frac{1}{4\pi^2}
\left\{\gamma_i(\mu)-\frac{1}{2}\frac{d\gamma_i(\mu)}{d\ln\mu}
\ln\left(\frac{M_{GUT}}{(M_{Pl}/m_{3/2})^{\alpha/2}\mu}\right)\right\}
\right.
\nonumber\\
&&
\left.\times
\ln\left(\frac{M_{GUT}}{(M_{Pl}/m_{3/2})^{\alpha/2}\mu}\right)
-\frac{1}{8\pi^2}Y_i\Big(\sum_jc_jY_j\Big)
g_Y^2(\mu)\ln\left(\frac{M_{GUT}}{\mu}\right)\right]
,
\eea
where $\gamma_i(\mu)$ and $g_Y^2(\mu)$
denote the running anomalous dimensions and the running $U(1)_Y$ gauge coupling
at $\mu$, respectively, and $Y_i$ is the $U(1)_Y$ hypercharge
of $Q_i$. Here again we ignored the off-diagonal parts of
$\omega_{ij}=\sum_{kl}y_{ikl}y^*_{jkl}$,
and the last part of $m_i^2(\mu)$
depending on $\sum_i c_iY_i$ arises as a consequence of
the ${\rm Tr}(Ym^2)$-term in the RG equation of $m_i^2$.
In generic situation,
the above results will be modified by the Yukawa couplings
for $a_{ijk}\neq 1$ or $c_i+c_j+c_k\neq 1$.
However for the first and second generations of quarks and leptons,
the modification will be negligible since the involved Yukawa couplings
are small enough.
Again the anomaly mediated contributions at $M_{GUT}$
leads to a mirage messenger scale
$(m_{3/2}/M_{Pl})^{\alpha/2}M_{GUT}$.
Also if $\alpha\approx 2$, the TeV scale
sfermion masses are approximately same as
the pure modulus-mediated sfermion masses without any RG running effect:
\bea
m_i^2(M_{SUSY})\,\approx\, c_i\left|M_0\right|^2
\quad \mbox{for} \quad
\alpha\approx 2
\eea
up to ignoring the corrections due to $y_{ijk}$
for $a_{ijk}\neq 1$ or $c_i+c_j+c_k\neq 1$.

The results of (\ref{lowgaugino}) and (\ref{lowsfermion})
show that the low energy soft parameters
in a mixed modulus-anomaly mediation with messenger scale
$\Lambda$ are approximately same
 as those
of the pure modulus-mediation with
a mirage messenger scale $(m_{3/2}/M_{Pl})^{\alpha/2}\Lambda$.
We note that this feature does not depend on the detailed form of
the modulus-mediation, thus applies to generic form of the mixed
modulus-anomaly mediation.
As was pointed out in \cite{choi2},
the mixed modulus-anomaly mediation can avoid
dangerous SUSY flavor and CP violations in a natural manner.
The soft terms  preserve the quark and lepton flavors if $n_i$ are flavor-independent,
which would arise automatically if the matter fields with common
gauge charge originate from the same geometric structure.
They also preserve CP since the relative CP phase between
$F^T$ and $F^C$ could be rotated away by the shift of the axion-component
of $T$ \cite{choi5}.
%
\begin{figure}[t]
\begin{center}
\begin{minipage}{15cm}
\centerline{
{\hspace*{-.2cm}\psfig{figure=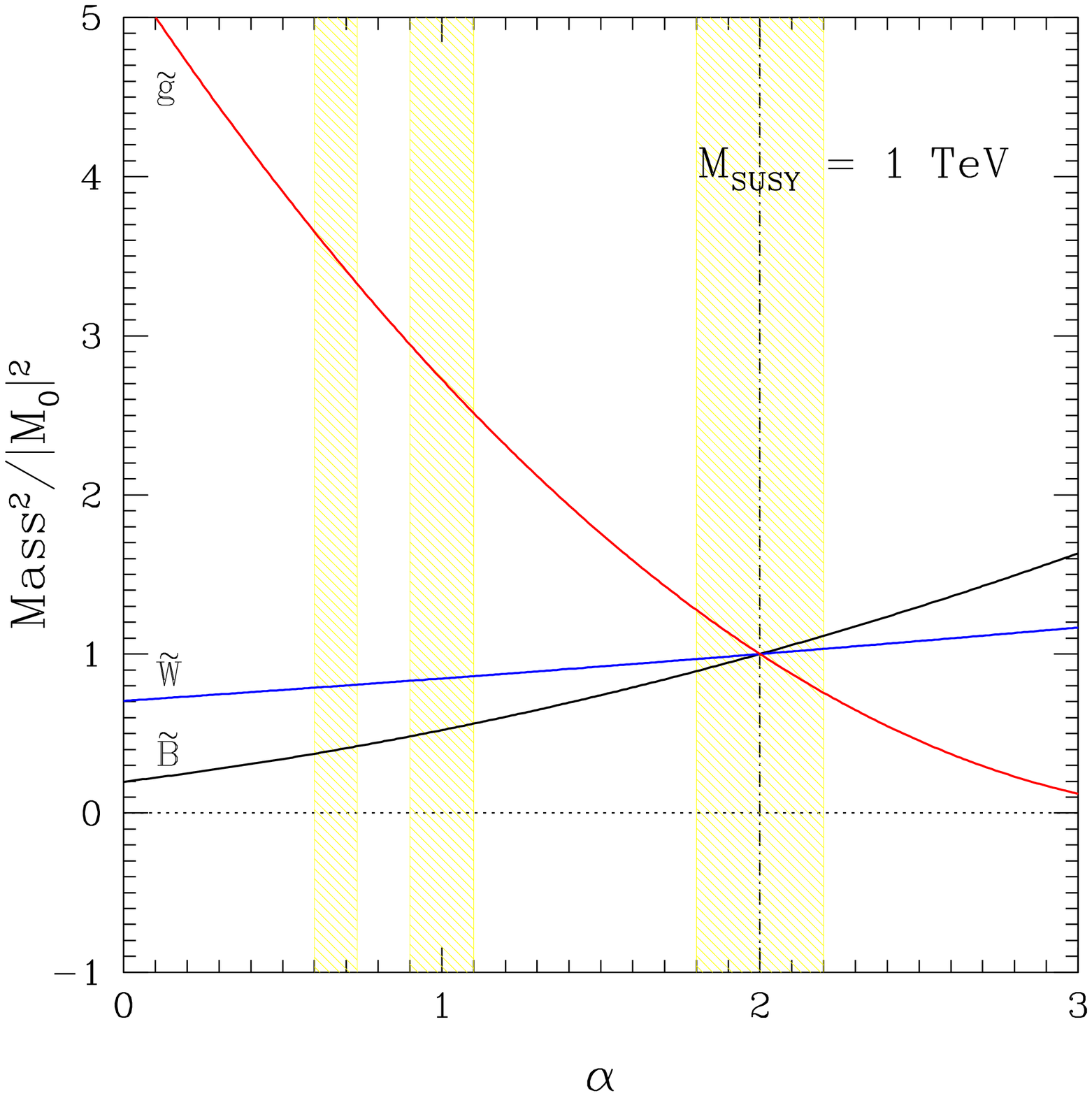,angle=0,width=7.5cm}}
{\hspace*{-.2cm}\psfig{figure=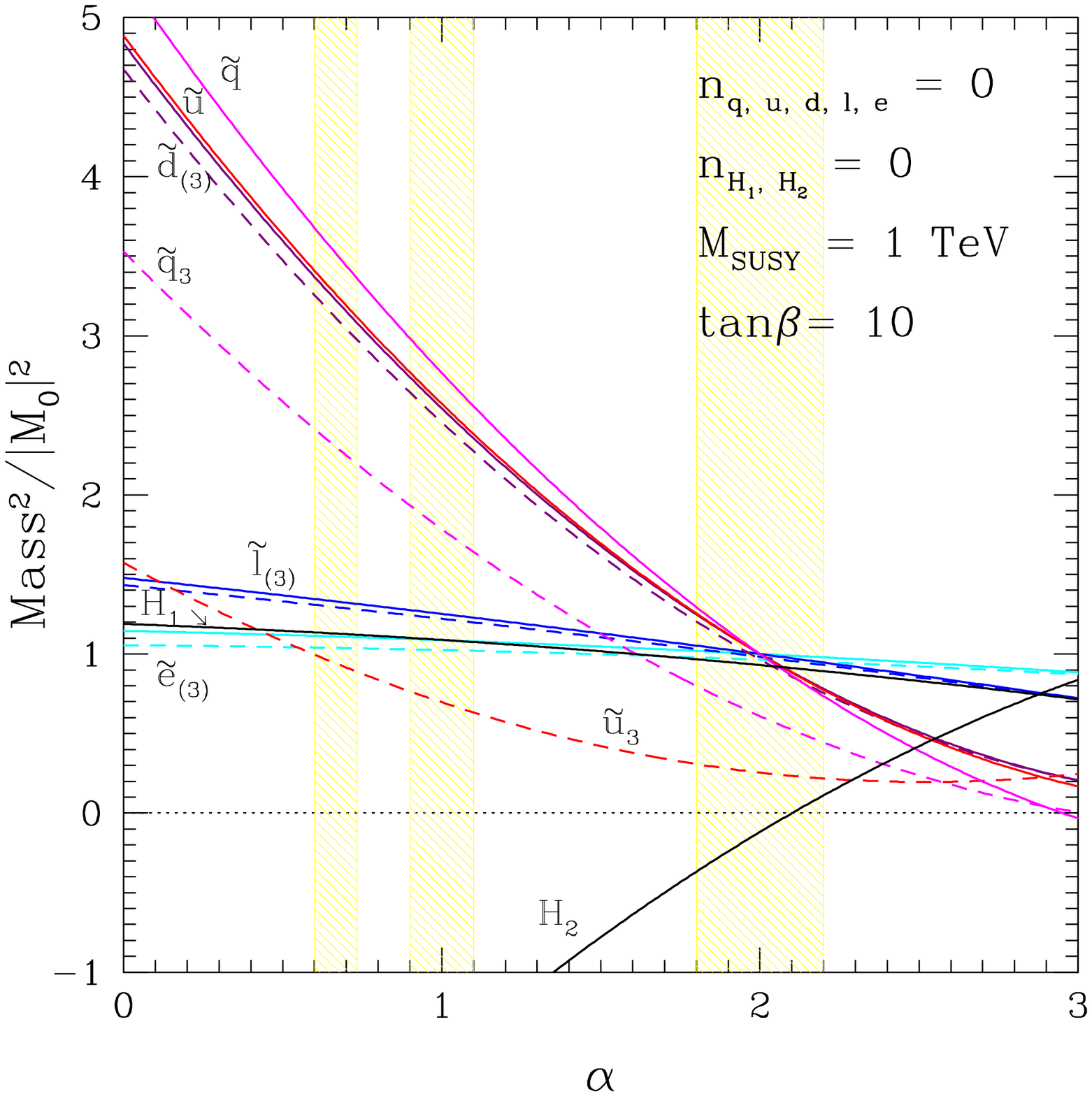,angle=0,width=7.5cm}}
}
\end{minipage}
%
\begin{minipage}{15cm}
\centerline{
{\hspace*{-.2cm}\psfig{figure=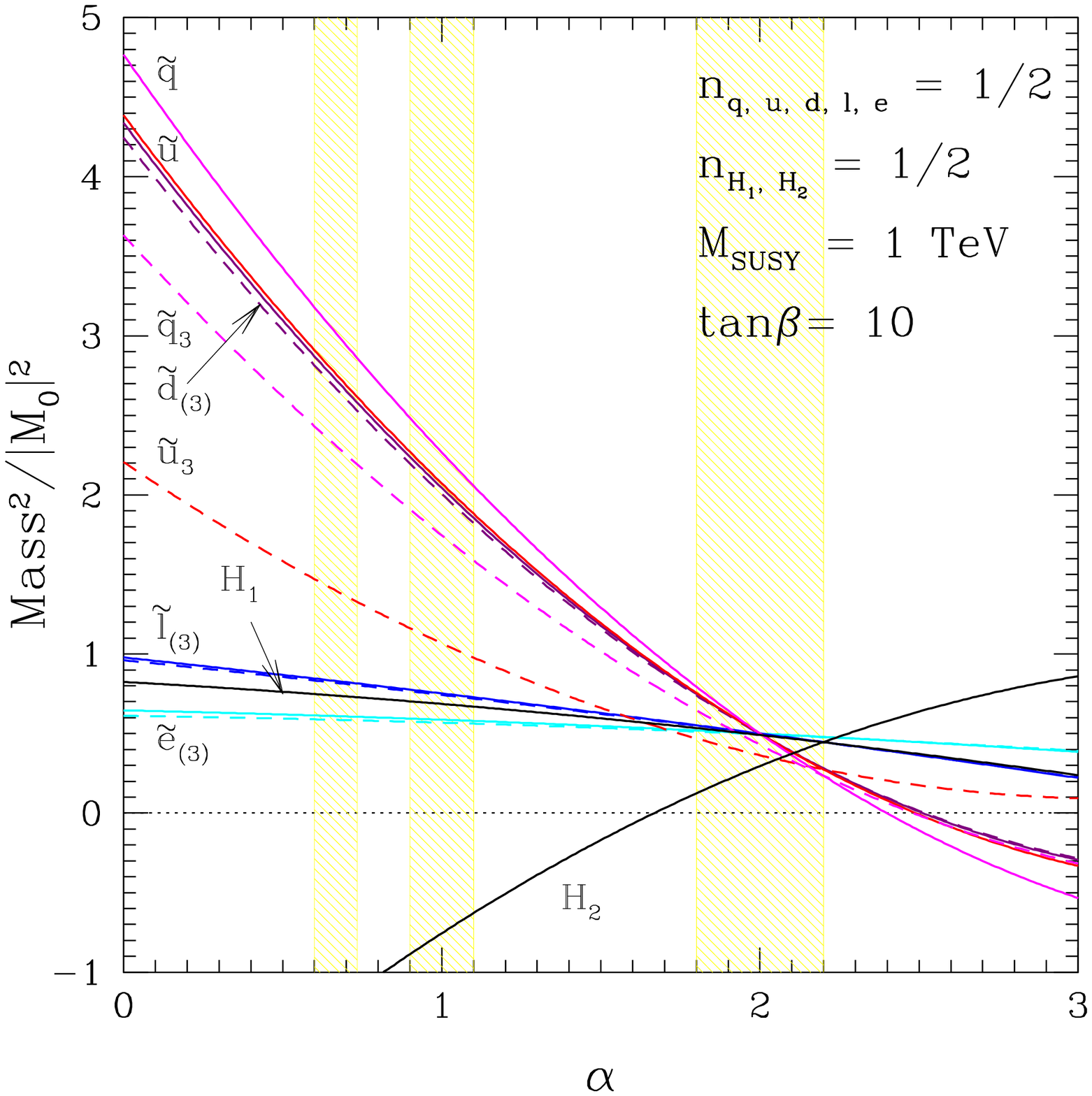,angle=0,width=7.5cm}}
{\hspace*{-.2cm}\psfig{figure=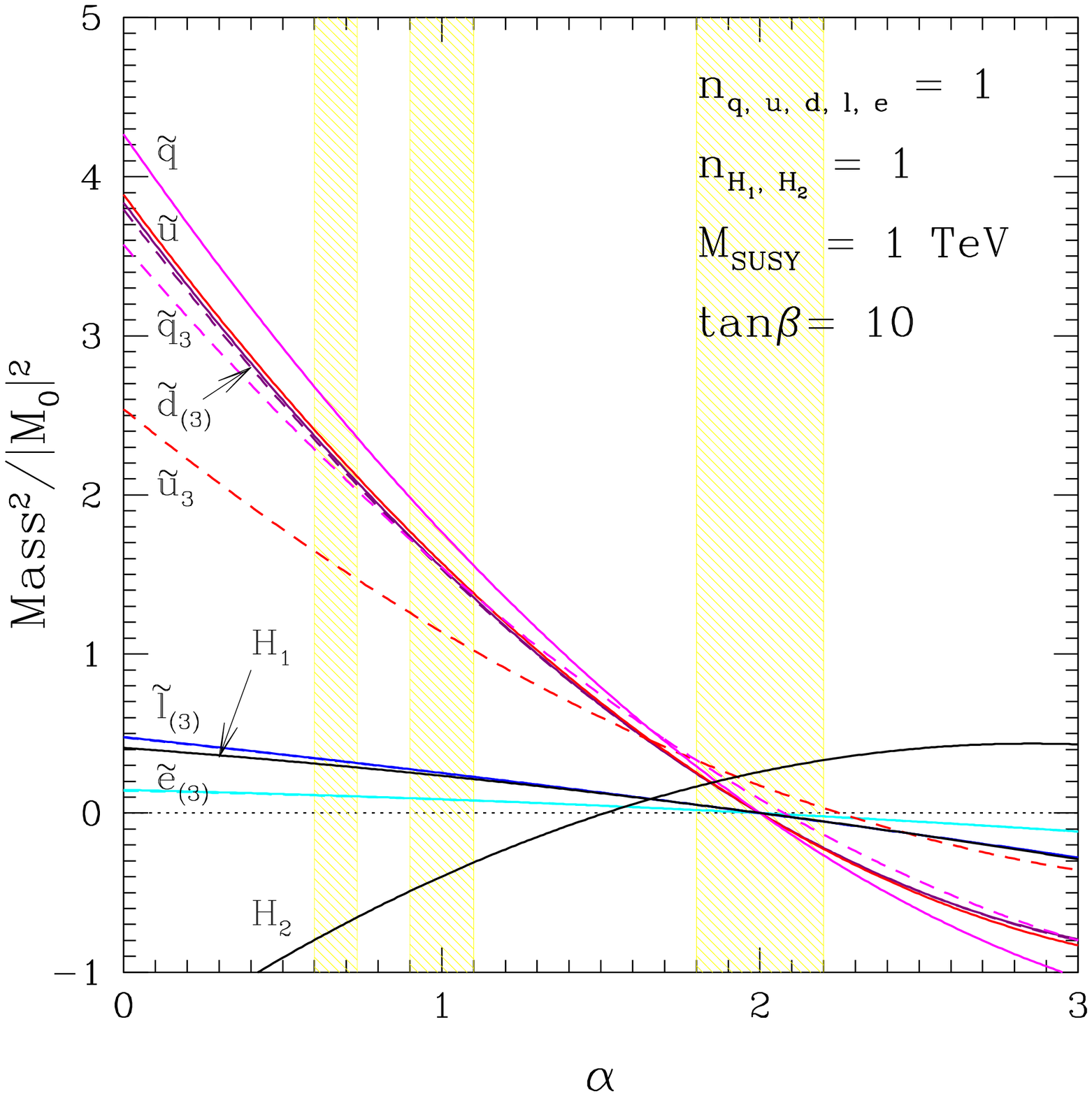,angle=0,width=7.5cm}}
}
\caption{
Sparticle mass spectrum at $M_{SUSY}=1$ TeV for
$0\leq \alpha\leq 3$.
The shaded regions correspond to the moduli K\"ahler and superpotential
(2.3) and the uplifting potential (2.13)
with $n_P=-1, 0, 1$ ($\alpha=2/3, 1, 2$),
 taking into account $10\%$ uncertainty.
Again the short-dashed curves denote the 3rd generation sfermions.
\label{fig:spectrum}
}
\end{minipage}
\end{center}
\end{figure}
%
%
In Figs. 1--\,3, we depict the pattern of low energy sparticle masses
for an appropriate range of $\alpha$.
In this calculation, we determined the gauge and Yukawa couplings
at $M_{SUSY}=1$ TeV using the 2-loop RG equations of the SM
and the top quark mass $m_t^{pole} = 178$  GeV.
Above $M_{SUSY}$, we have used the 1-loop RG equation
of the MSSM to arrive at $M_{GUT}$.
The results are not affected significantly
by a different choice of $M_{SUSY}$ as long as it is  not far from the weak scale.
Fig. 1 provides an overall view of the low energy sparticle spectrum
for the whole range of $\tan^{-1}(\alpha/4)$,
including the pure modulus mediation ($\tan^{-1}(\alpha/4)=0$)
and also the pure anomaly mediation ($\tan^{-1}(\alpha/4)=\pi/2$).
Here we assumed that all MSSM matter and Higgs fields originate from
$D7$ branes, thus $n_i=0$ for all MSSM chiral multiplets, and used the
top quark Yukawa coupling for $\tan\beta =10$.
Depending upon the value of $\alpha$,
the gluino to Wino mass ratio (and also
the squark to slepton mass ratios) can be
considerably suppressed (e.g. $\tan^{-1}(\alpha/4)\sim 0.22\pi$)
or enhanced (e.g. $\tan^{-1}(\alpha/4)\sim 0.75 \pi$), relative to the pure modulus
 mediation case.
Note that the range $0.20\pi \leq \tan^{-1}(\alpha/4)\leq 0.64\pi$ gives
either a tachyonic squark or a tachyoinc slepton.
For $0.64\pi \leq \tan^{-1}(\alpha/4)\leq 0.77\pi$, the negative slepton mass-square
of the pure anomaly mediation can be lifted to a positive value,
while keeping the Wino LSP.
The relative signs of Bino and gluino masses against the Wino mass
 can be $(+,+)$ or $(-,+)$ since the $(+,-)$ case is excluded due to a tachyonic
slepton/squark.

Fig. 2 shows the low energy sparticle  masses for $0\leq \alpha \leq 3$.
This range of $\alpha$ contains  $\alpha=1$ predicted by the minimal KKLT set-up
and also $\alpha=2$ which has a mirage messenger scale close to
TeV. As for the squark/slepton spectrums, we considered three different cases
distinguished by the universal values of $n_i$:
$0, 1/2$ or $1$.
For $\alpha= 1$ which is predicted by the minimal KKLT set up, we summarized the
resulting sparticle spectrums at $M_{SUSY}$ in Table. 1.
Note that the gluino to Wino/Bino mass ratios
$M_{\tilde{g}}/M_{\tilde{W},\tilde{B}}$
and also the squark to slepton mass ratios $m_{\tilde{q}}/m_{\tilde{l}}$
for the minimal KKLT value
are significantly smaller than the values predicted by  the general mSUGRA scenario
or the pure anomaly mediation scenario.
For $\alpha= 2$, the low energy superparticle masses are approximately unified
(up to corrections due to the top quark Yukawa coupling)
since it gives a mirage messenger scale close to 1 TeV.
Note that for a fixed value of $\alpha$,
the sparticle mass ratios discussed here
are  insensitive to the overall size of the
SUSY breaking order parameter $F^T/T$ as well as of
the details of the electroweak symmetry breaking
which will be discussed below.

\begin{table}[h]
\begin{tabular}{|c|c|c|c|}
\hline
  & $\alpha=1$
& $(\alpha=1)/(\alpha=0)$ \\
\hline
$\tilde{B}$ & 0.79 & 1.48\\
\hline
$\tilde{g}$ & 1.79 & 0.66\\
\hline
\end{tabular}

\vspace{1em}
\begin{tabular}{|c|c|c|c|c|c|c|}
\hline
& \multicolumn{6}{|c|}{ $\tan\beta = 10$}\\
\hline
&
\multicolumn{2}{|c|}{$n_i = 0$}
&
\multicolumn{2}{|c|}{$n_i = 1/2$}
&
\multicolumn{2}{|c|}{$n_i = 1$}\\
\hline
& $\alpha = 1$  & $(\alpha =1)/(\alpha = 0)$ & $\alpha = 1$
 &$(\alpha=1)/(\alpha=0)$& $\alpha = 1$ & $(\alpha = 1)/(\alpha = 0)$\\
\hline
$\tilde{e}$      & 1.13 & 0.89 & 0.83 & 0.87 & 0.32 & 0.71 \\
\hline
$\tilde{e}_3$    & 1.10 & 0.90 & 0.82 & 0.88 & 0.32 & 0.71 \\
\hline
$\tilde{\ell}$   & 1.22 & 0.84 & 0.94 & 0.80 & 0.55 & 0.66 \\
\hline
$\tilde{\ell}_3$ & 1.20 & 0.84 & 0.94 & 0.80 & 0.55 & 0.66 \\
\hline
$\tilde{d}$      & 1.73 & 0.66 & 1.55  & 0.63 & 1.35  & 0.58 \\
\hline
$\tilde{d}_3$    & 1.70 & 0.66 & 1.54  & 0.63 & 1.35 & 0.58 \\
\hline
$\tilde{u}$      & 1.74 & 0.66 & 1.56  & 0.63 & 1.36  & 0.58 \\
\hline
$\tilde{u}_3$    & 0.91 & 0.61 & 1.12  & 0.64 & 1.16  & 0.61 \\
\hline
$\tilde{q}$      & 1.81 & 0.66 & 1.64  & 0.63 & 1.44  & 0.59 \\
\hline
$\tilde{q}_3$    & 1.45 & 0.65 & 1.44  & 0.63 & 1.35  & 0.60 \\
\hline
\end{tabular}
\caption{Sparticle spectrum of the minimal KKLT set-up ($\alpha=1$)
at $M_{SUSY}$. All masses are divided by the Wino mass $M_2$ at 1 TeV.
The ratios to the pure modulus mediation ($\alpha=0$) are also presented.}
\end{table}

Let us now examine the electroweak symmetry breaking
in the mixed modulus-anomaly mediation scenario.
This issue depends  on how the Higgs mass parameters
$\mu$ and $B$ are generated.
An important phenomenological issue which is sensitive to
$\mu$ is the nature of the lightest supersymmetric particle (LSP).
Also if one could have a concrete theoretical scheme to relate the $\mu$ and $B$
to $M_0=F^T/(T+T^*)$,  the overall size of $M_0$ might be constrained
by the condition of the correct electroweak symmetry breaking.
Here we will restrict the analysis to the  simplest (though not the most attractive) scheme to generate
$\mu$ and $B$: the minimal supersymmetric standard model (MSSM) with
$\mu$ and $B$ obtained from  (\ref{higgsbilinear}) or (\ref{kkltmu})
under appropriate fine-tuning,
while leaving the analysis for the next minimal supersymmetric standard model (NMSSM)
with a singlet $N$, i.e. the model of (\ref{nmssm}), for future work.
We also limit the analysis to the tree-level
Higgs potential  for simplicity.

\begin{figure}[t]
\begin{center}
\begin{minipage}{15cm}
\centerline{
{\hspace*{-.2cm}\psfig{figure=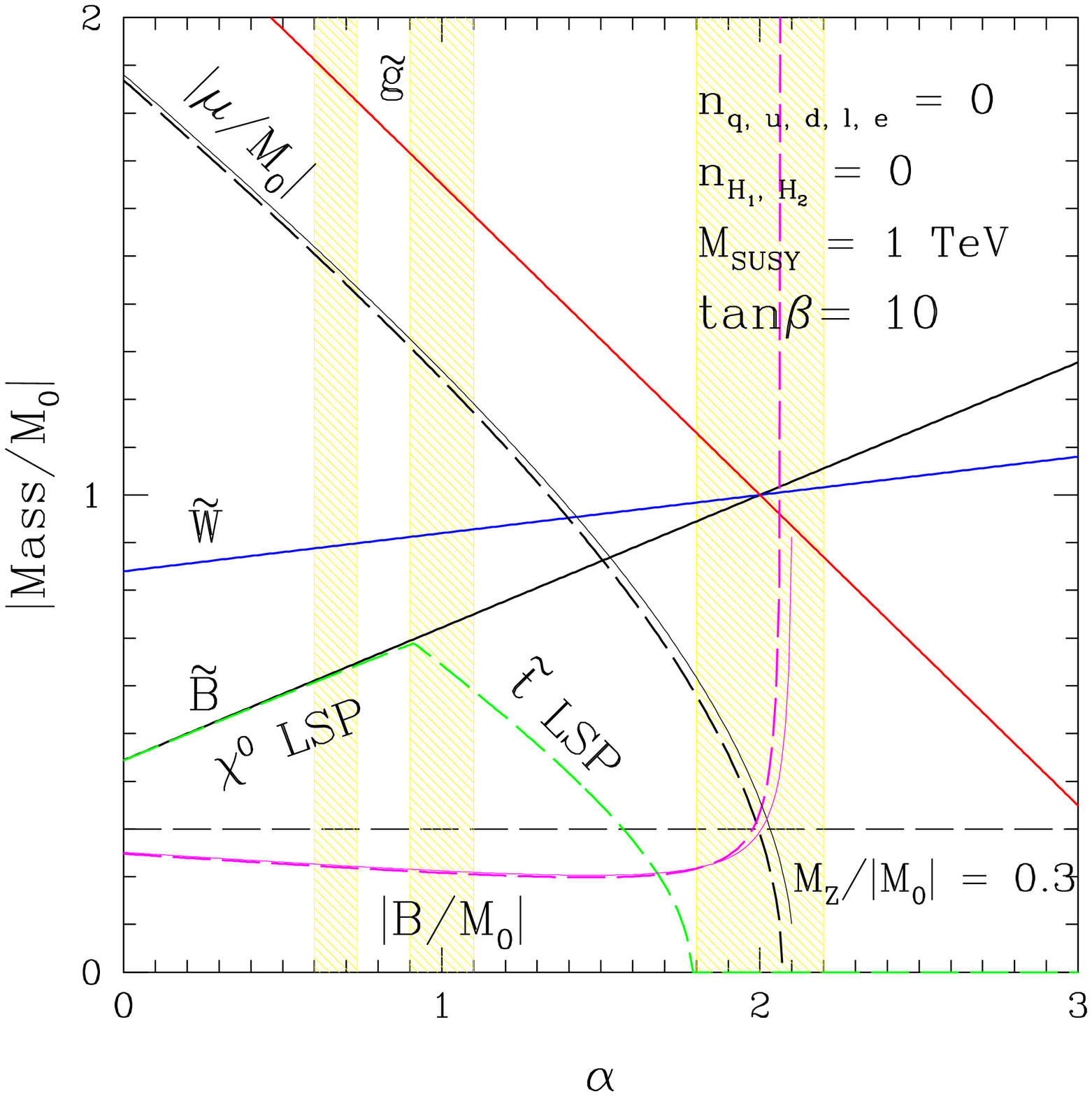,angle=0,width=5.0cm}}
{\hspace*{-.2cm}\psfig{figure=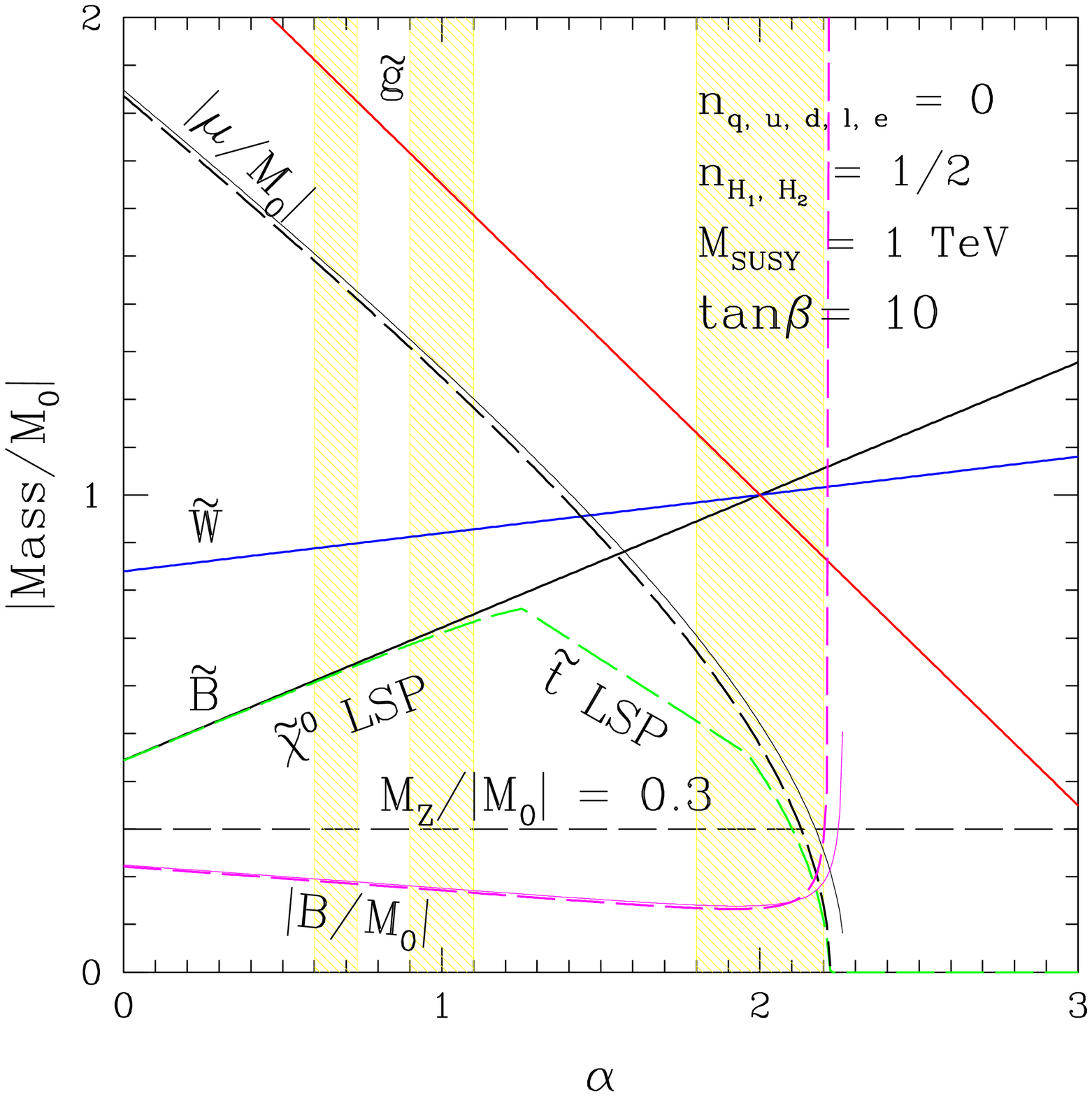,angle=0,width=5.0cm}}
{\hspace*{-.2cm}\psfig{figure=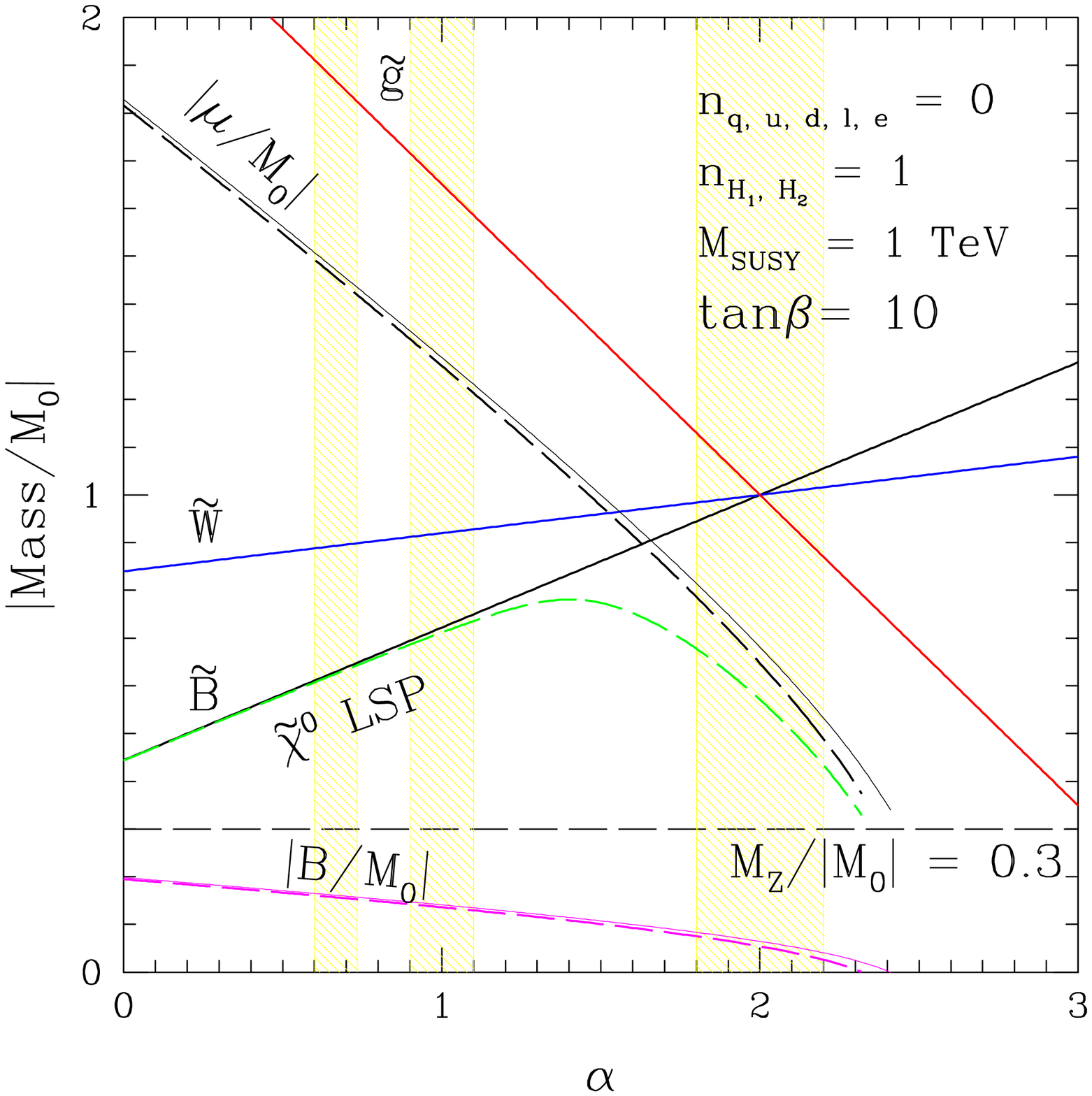,angle=0,width=5.0cm}}
}
\centerline{
{\hspace*{-.2cm}\psfig{figure=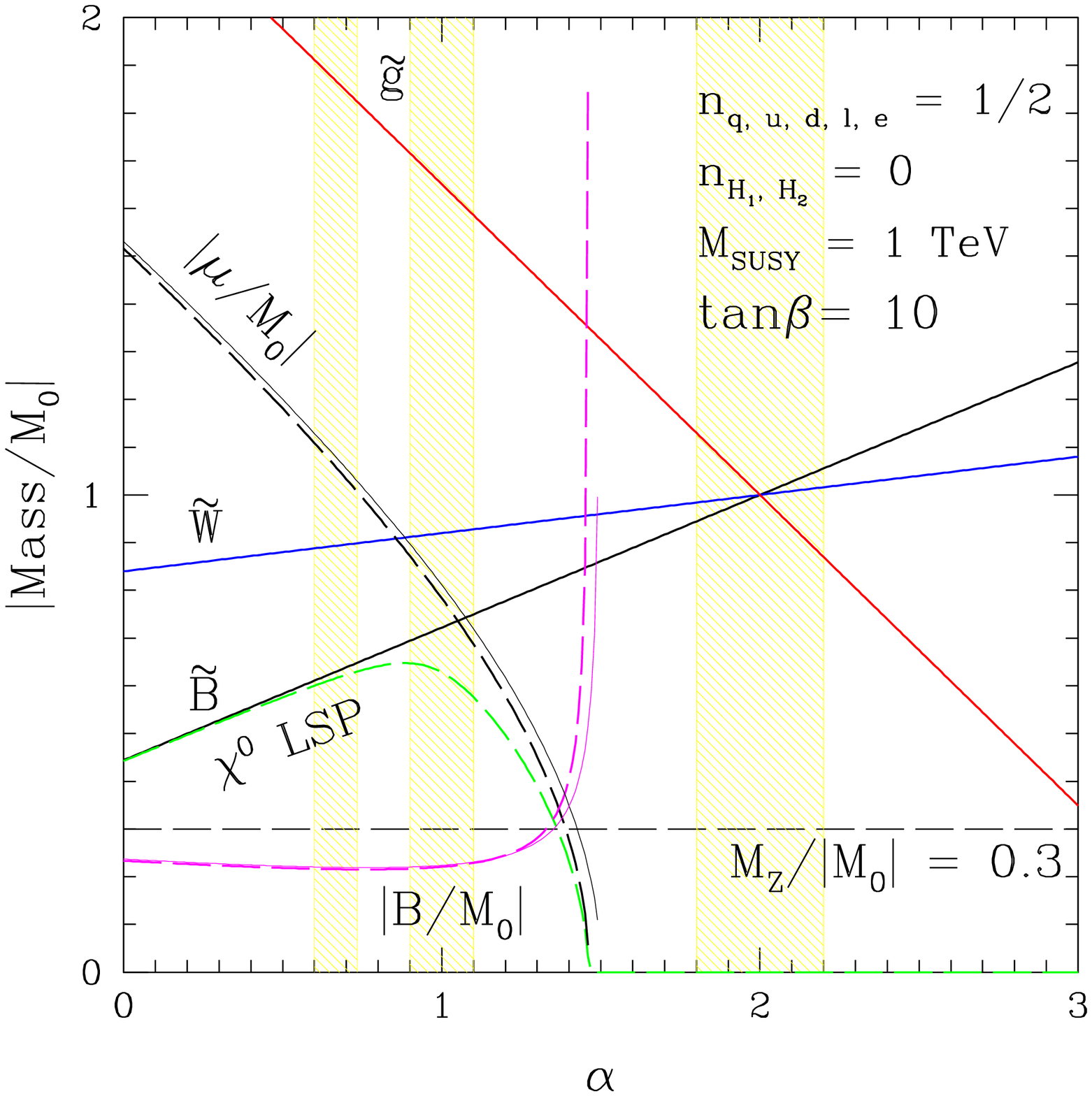,angle=0,width=5.0cm}}
{\hspace*{-.2cm}\psfig{figure=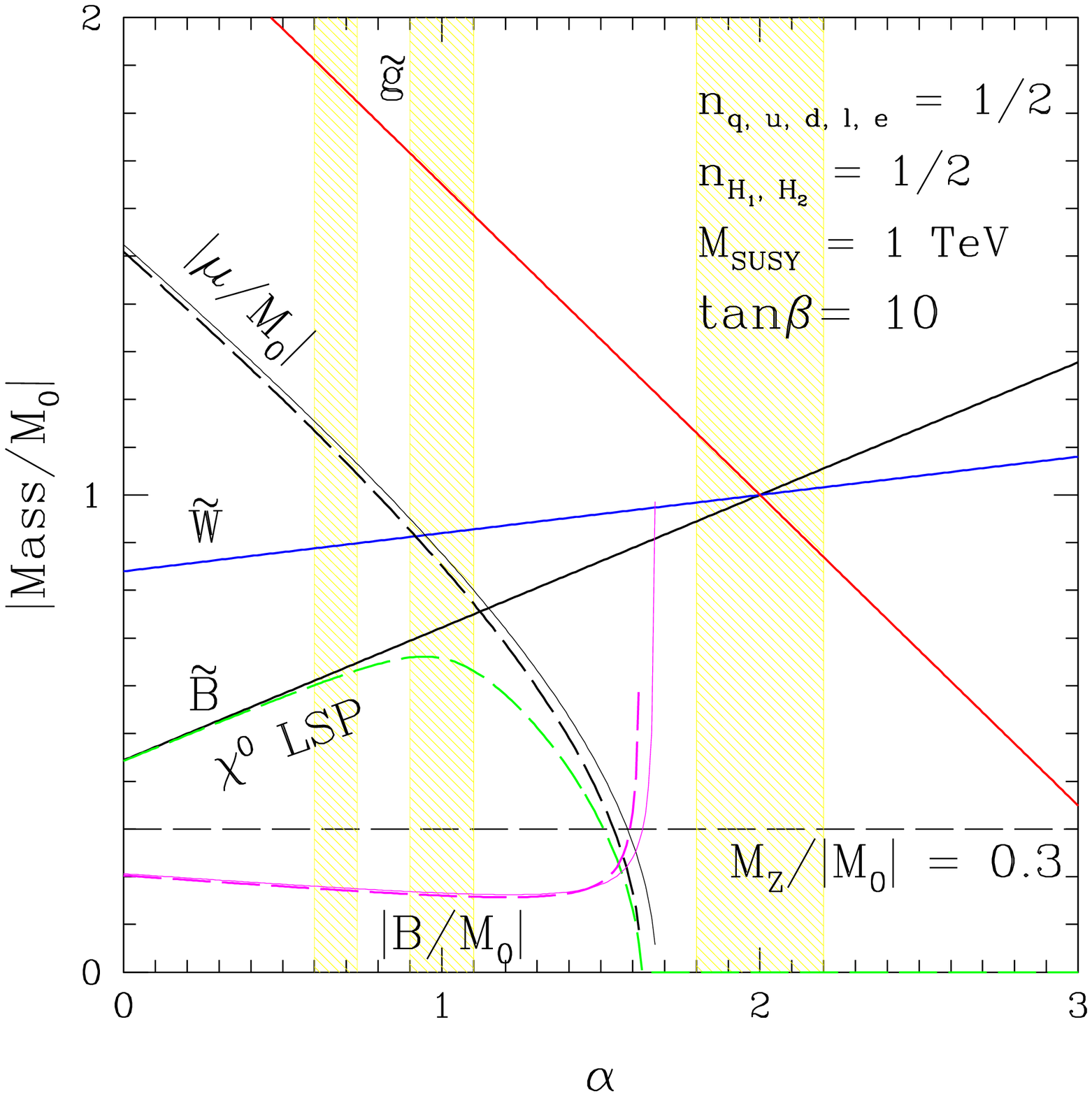,angle=0,width=5.0cm}}
{\hspace*{-.2cm}\psfig{figure=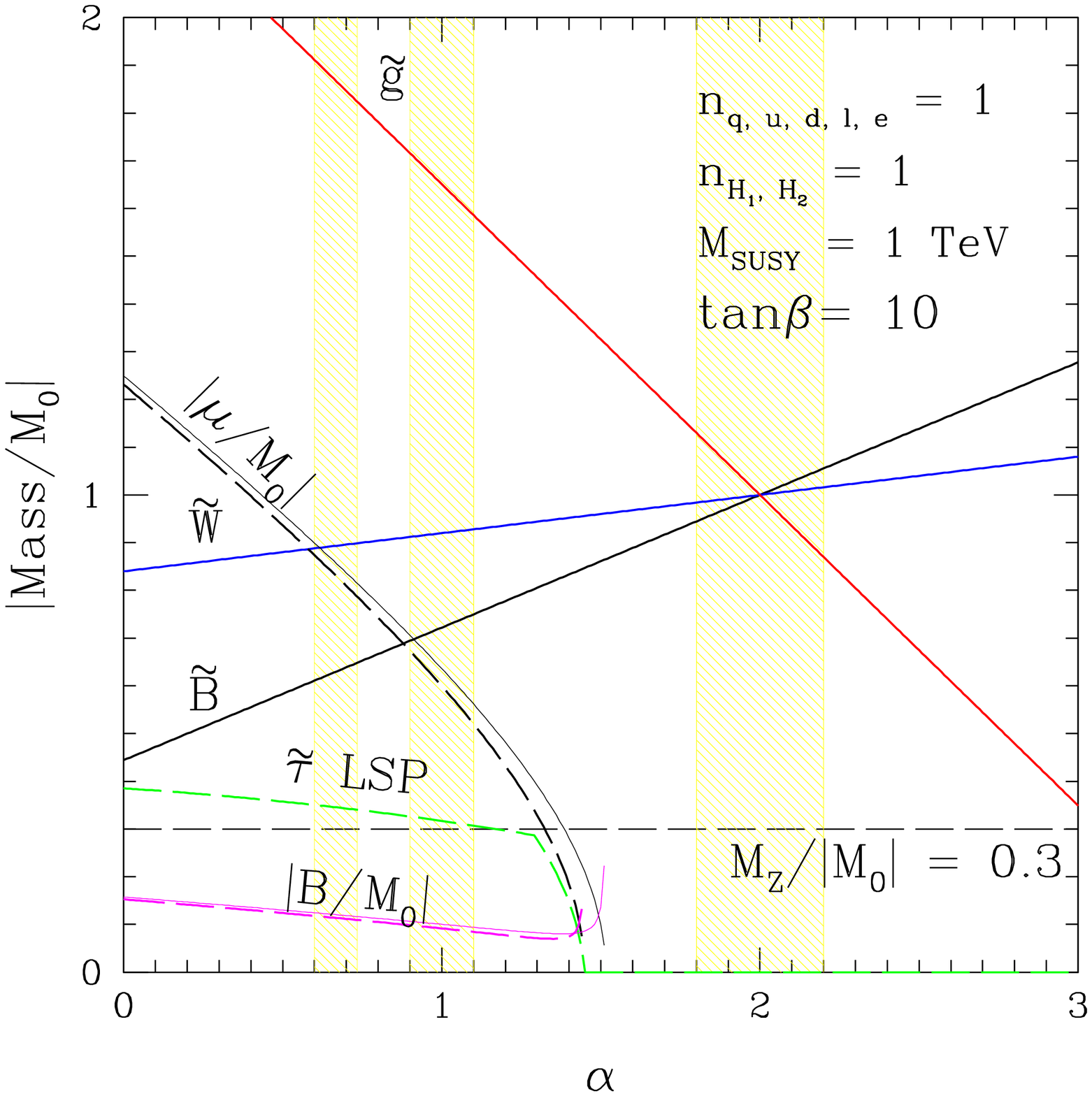,angle=0,width=5.0cm}}
}
\caption{The behavior of the Higgsino mass parameter $\mu$.
The shaded region is same as in Fig.2.
The dashed (thin--solid) curve corresponds to
$M_Z=0.3 M_0$ ($M_0/M_Z\rightarrow \infty$).
\label{fig:rewsb_mssm}}
\end{minipage}
\end{center}
\end{figure}

The neutral part of the Higgs potential is given by
\beq
V=\tilde{m}_1^2|H_1^0|^2+\tilde{m}_2^2|H_2^0|^2-\l(
B\mu H_1^0H_2^0+\mbox{c.c.}\r)+
\frac{1}{8}(g_1^2+g_2^2)\l(|H_1^0|^2-|H_2^0|^2\r)^2\,,
\eeq
where $\tilde{m}_{1,2}^2=m_{H_{1,2}}^2+|\mu|^2$.
If the Higgs soft masses and $\mu$ satisfy
the conditions for a symmetry breaking stable vacuum \cite{Inoue:1982pi}:
\bea
\tilde{m}_1^2 \tilde{m}_2^2 - |B\mu|^2 < 0 &,&~~~
\tilde{m}_1^2 + \tilde{m}_2^2 - 2 |B\mu| > 0,
\eea
we obtain the following relations:
\bea
\label{eq:rewsb_condition}
\mu^2 = -\frac{M_Z^2}{2} +\frac{m^2_{H_1}-m^2_{H_2}\tan^2\beta}{\tan^2\beta-1}
&,&~~~|B\mu| = \frac{\tan\beta}{1+\tan^2\beta}(m^2_{H_1}+m^2_{H_2}+2\mu^2),
\eea
which allow us to determine $\mu/M_0$ and $B/M_0$ in terms of $M_Z/M_0$, $\tan\beta$ and
$m_{H_i}/M_0$.
In Fig.\ref{fig:rewsb_mssm}, we plot the resulting $\mu/M_0$, $B/M_0$ and the LSP mass
for  various choices of $n_i$ of the squarks, sleptons and Higgs
fields which are assumed to have the K\"ahler metric $Z_i=(T+T^*)^{-n_i}$.
As is shown, the qualitative behavior of $\mu$ and $B$ is common to all
 models, while the precise position of the
 curves and the nature of LSP are sensitive to the choice of $n_i$.
Thin sold curve for $\mu$ indicates the value of $\mu$ in the limit
$M_Z/M_0\rightarrow 0$, while the dashed curve represents
$\mu$ for $M_Z/M_0=0.3$.
Positiveness of $M_Z^2$
 in (\ref{eq:rewsb_condition}) ensures that there is no symmetry breaking solution
 for $|\mu|$ above the thin solid curve.
Because the slepton masses are typically  ${\cal O}(M_0)$,
 we are required to choose $M_0$ significantly bigger than $M_Z$.
Note that $\mu/M_0$ is
almost independent of $M_Z/M_0$ for $M_0\geq 3.3M_Z$,
which amounts to the well known fine-tuning of $\mu$ required for
the correct electroweak symmetry breaking in MSSM
 \cite{Barbieri:1987fn}.
Starting from the pure modulus mediation ($\alpha=0$), increasing
the anomaly mediated contribution eventually erases
the $y_t$-induced radiative correction to $m^2_{H_2}$
 and finally
restores the electroweak gauge symmetry, which corresponds to
the value of $\alpha$ for which $\mu=0$.
The value of  $|B|$ blows up at this value of $\alpha$ unless
$|B|$ becomes zero before arriving this value of $\alpha$,
which would make  the Higgs potential unbounded from below.
These features can be understood by noting that
the mirage messenger scale is given by $(m_{3/2}/M_{Pl})^{\alpha/2}M_{GUT}$,
thus the strength of radiative electroweak symmetry breaking
\cite{Inoue:1982pi,rewsb}
becomes weaker
when $\alpha$ increases from 0 to a positive value of ${\cal O}(1)$.

When $\alpha$ increases from 0 to a positive value,  the lightest neutralino is changed from
the Bino-like to the Higgsino-like around the point where $\mu$ crosses the Bino mass.
Typically, the minimal KKLT value ($\alpha = 1$) corresponds to the Bino or Bino--Higgsino
mixing region while $\alpha = 2$ corresponds to an almost pure Higgsino,
which will lead to a considerably different consequence in the dark matter
 scenario relative to the general mSUGRA case if the mixing is
 sufficiently large ($\alpha=1$) or Higgsino is not too light
 ($\alpha=2$) \cite{darkmatter}.
The model-dependence of the value of $\mu$
originates mainly from
the top Yukawa  coupling.
On the other hand, $\mu$ is rather insensitive to $\tan\beta$
 except for $\tan\beta \simeq 1$ which is disfavored by the
 Higgs boson search in LEPII \cite{Eidelman:2004wy}. This is because most of
 the contribution to $\mu^2$ in (\ref{eq:rewsb_condition}) comes from $m^2_{H_2}$
 as the contribution from   $m^2_{H_1}$ is strongly suppressed by $\tan^{-2}\beta$.
 The green curves in Fig.\ref{fig:rewsb_mssm} indicate
 the LSP mass in the unit of $M_0$.
The nature of LSP is somewhat model-dependent,
 and can be neutralino, stau or stop, depending upon
the choice of $n_i$, $\alpha$ and $\tan\beta$.
If the LSP is stable, this feature can provide a strong constraint on the model.
In general, a heavier $H_2$ lowers the stop mass through
the radiative correction involving the top Yukawa coupling,
while the choice $n_i=1$  or a large $\tan\beta$
 makes the slepton lighter.

\begin{figure}[t]
\begin{center}
\begin{minipage}{15cm}
\centerline{
{\hspace*{-.2cm}\psfig{figure=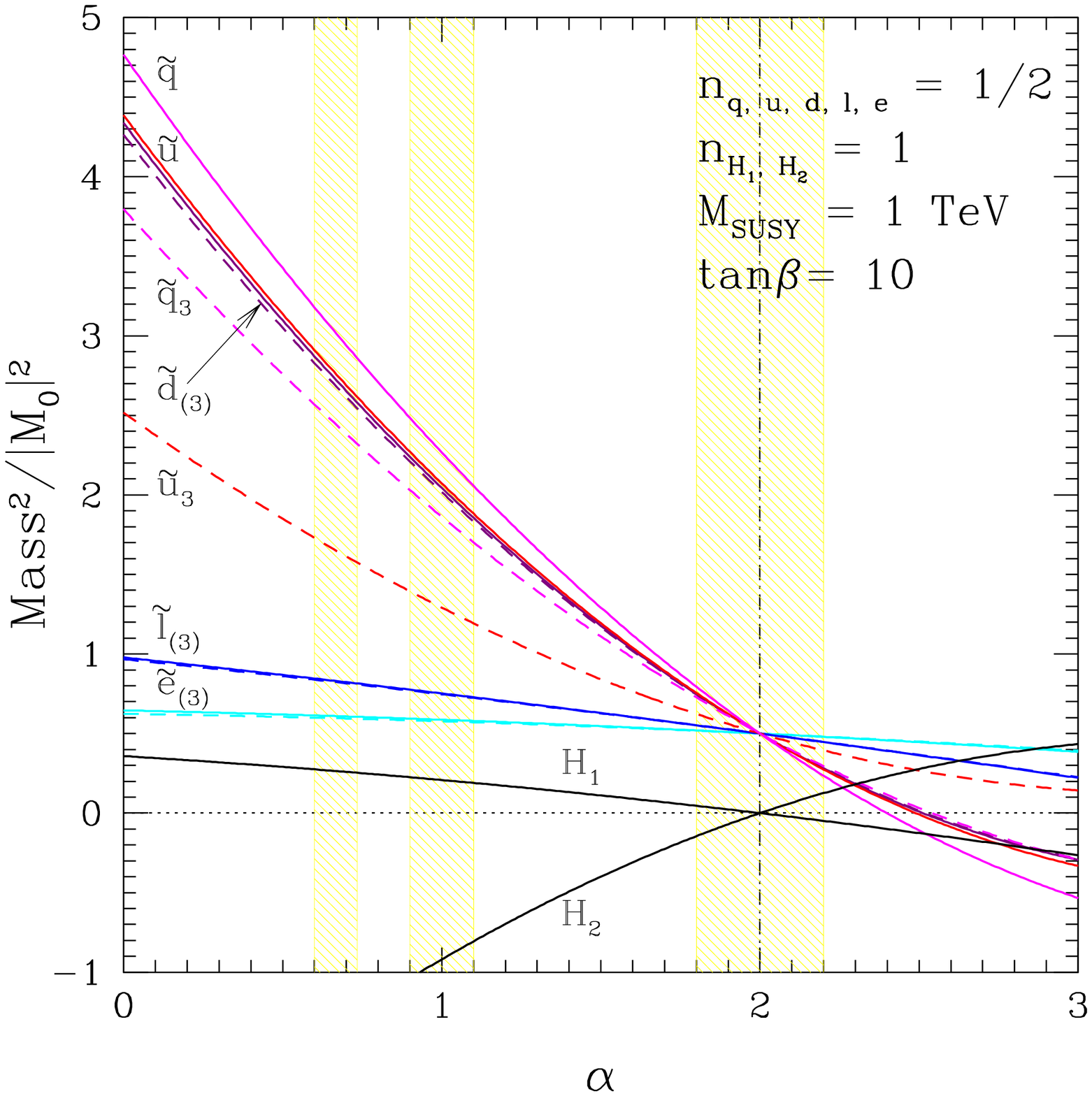,angle=0,width=7.5cm}}
{\hspace*{-.2cm}\psfig{figure=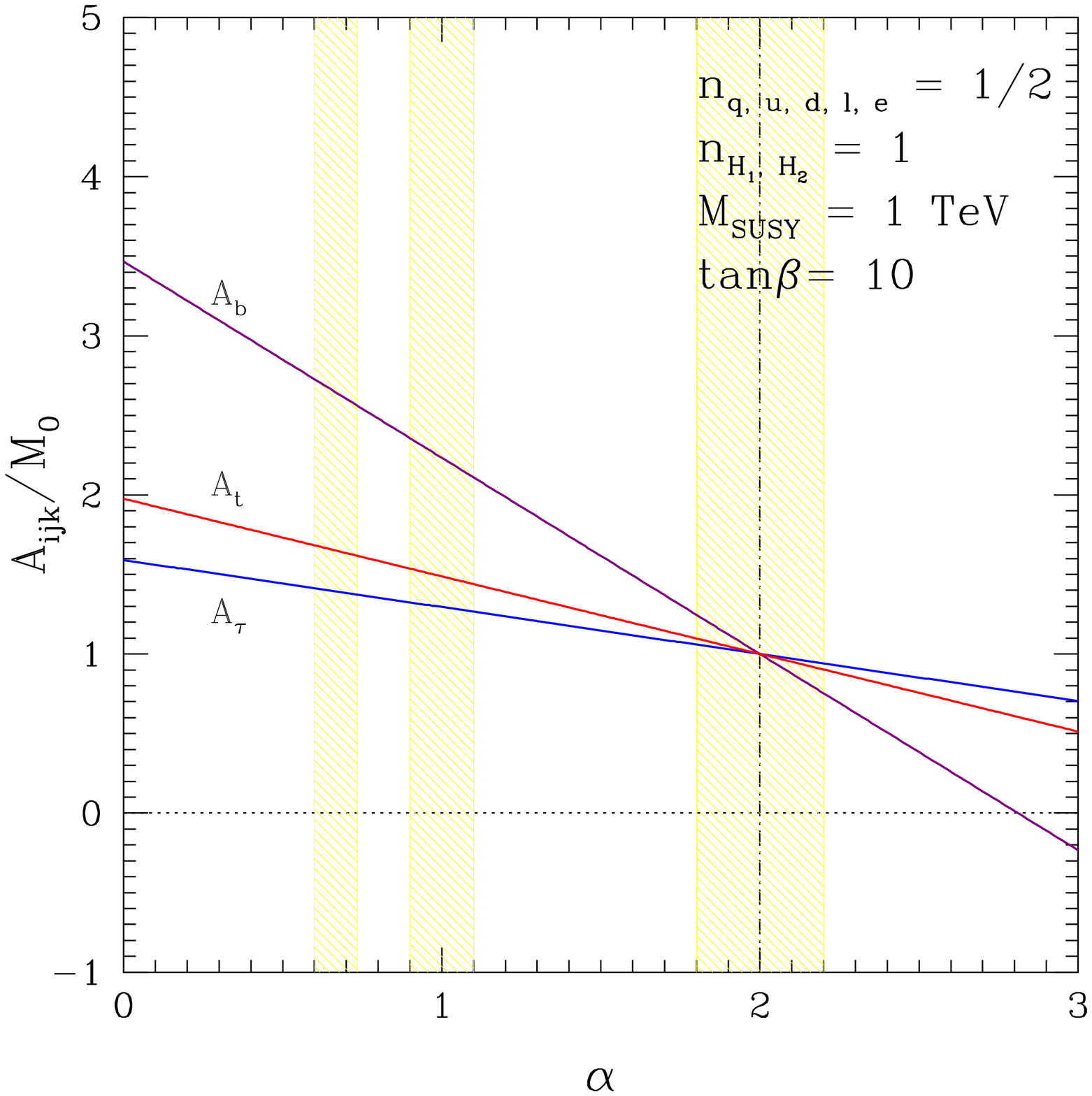,angle=0,width=7.5cm}}
}
\centerline{
{\hspace*{-.2cm}\psfig{figure=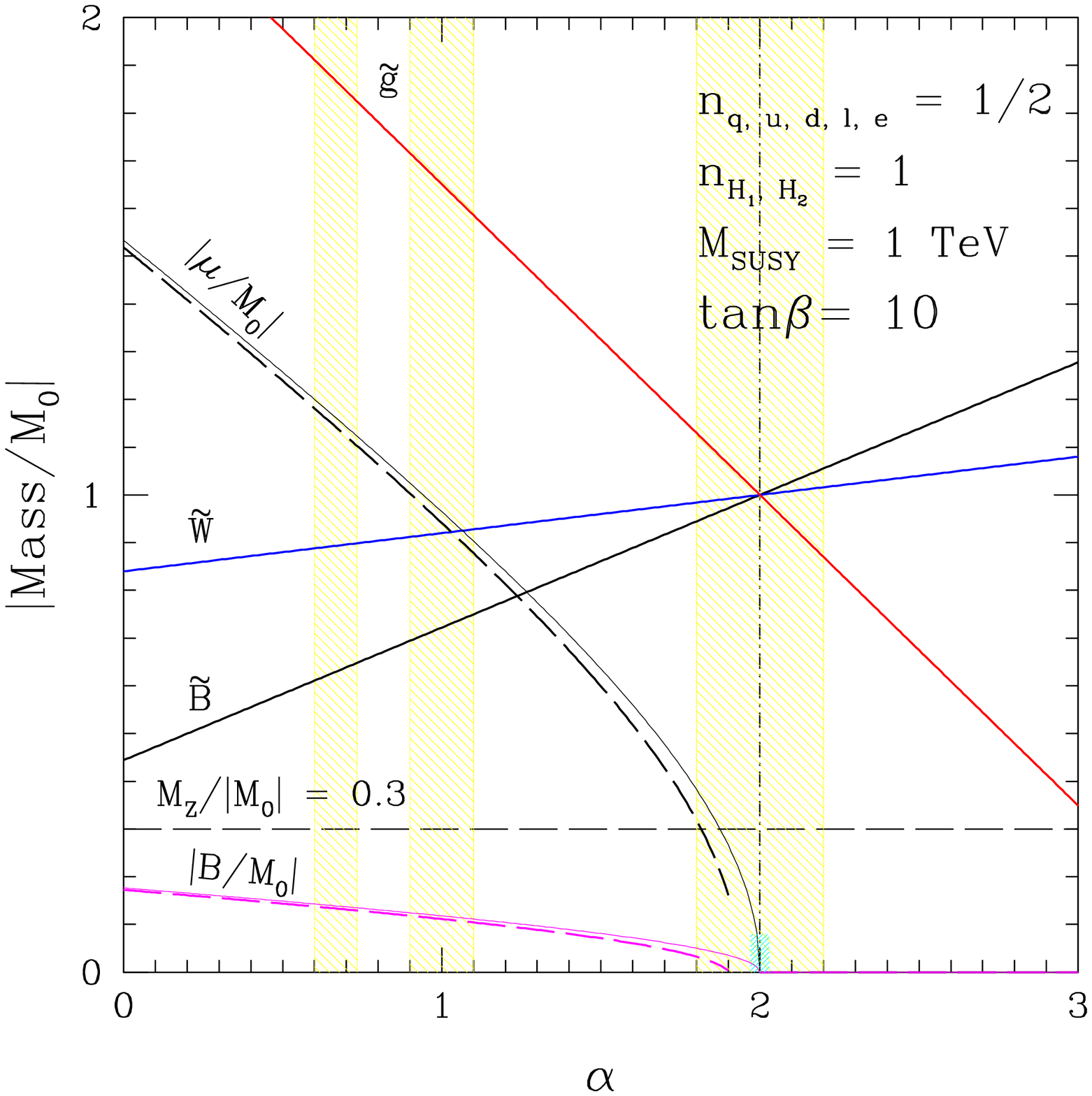,angle=0,width=7.5cm}}
}
\caption{A model for unified soft parameters at $M_{SUSY}$
for $\alpha=2$.
We choose $n_{q,u,d,l,e}=1/2$ and $n_{H_1, H_2}=1$ to obtain
$a_{ijk}=c_i+c_j+c_k=1$.
A little hierarchy between the Higgs boson masses
and the sparticle masses for $\alpha=2$ is protected at 1-loop level.
Note that the results on $\mu/M_0$ and $B/M_0$ at $\alpha\approx 2$
have an uncertainty of ${\cal O}(10^{-1})$ due to the threshold corrections
of ${\cal O}(M_0^2/8\pi^2)$ to the Higgs mass-squares at $M_{GUT}$.
\label{fig:hierarchy}}
\end{minipage}
\end{center}
\end{figure}

As we have noted, if $y_{ijk}\neq 0$ only for $a_{ijk}=1$ and
$c_i+c_j+c_k=1$ which would be obtained when $n_i+n_j+n_k=2$,
and also if $\alpha\approx 2$,
the RG evolution of modulus mediation is almost canceled
by the anomaly mediation, which results in the mirage messenger scale
close to TeV.
This set-up may provide
 a new insight for the little
 hierarchy problem in supersymmetric standard model.
In Fig.\ref{fig:hierarchy}, we choose $n_i=1/2$ for matter fields and
 $n_i=1$ for the Higgs fields, satisfying the condition
that $n_i+n_j+n_k=2$ for nonzero Yukawa couplings.
The modulus-mediated soft mass-squares of the Higgs bosons
at $M_{GUT}$ are vanishing up to small threshold corrections of ${\cal O}(M_0^2/8\pi^2)$,
and thus $m_{H_i}^2(M_{SUSY})/M_0^2={\cal O}(1/8\pi^2)$  for $\alpha=2$.
One the other hand, the modulus-mediated squark/slepton mass squares and the
modulus-mediated gaugino mass at $M_{SUSY}$
are  $M_0^2/2$ and $M_0$, respectively, thus
$M_a(M_{SUSY})\approx M_0$ and $m^2_{\tilde{q},\tilde{l}}
(M_{SUSY})\approx M_0^2/2$ for $\alpha=2$.
This might enable us to generate  a little hierarchy between
the weak scale and the sparticle mass in a natural manner:
$m_H^2/M_0^2 ={\cal O}(1/8\pi^2)$.
Of course, we then need a mechanism
to generate $\mu$ and $B$ smaller than $M_0$
by one order of magnitude.
In general, keeping the hierarchy of ${\cal O}(10^{-2})$
between the Higgs mass-squares
 and  the gaugino/squark mass-squares is highly non-trivial because
 the latter enters in the former through the radiative corrections involving
the strong coupling constants \cite{Casas:2003jx}.
However, in the mixed modulus-anomaly mediation with $\alpha\approx 2$,
those radiative corrections are automatically canceled by the anomaly-mediated
contributions. Note that although no string theory realization is found yet,
$\alpha=2$
can be naturally obtained by an effective theory with
uplifting potential $V_{\rm lift}\propto 1/(T+T^*)$.
%
%
%
%

\section{Conclusion}

In this paper, we have examined some phenomenological consequences of
the mixed modulus-anomaly mediation scenario for supersymmetry breaking
in which the modulus mediation and the anomaly mediation give comparable
contributions to soft parameters at the messenger scale
$\sim M_{GUT}$.
Such mediation scheme  can arise naturally in compactified string
and  brane models which stabilize the light moduli by nonperturbative effects
at a supersymmetric AdS vacuum and then break SUSY by a sequestered uplifting
potential.
A concrete example of such scenario has been proposed recently by KKLT \cite{Kachru:2003aw}
and the pattern of resulting soft terms are analyzed in \cite{choi1,choi2}.
The scheme may also offer an interesting cosmological scenario
which produces the correct amount of the neutralino dark matter while
avoiding the cosmological moduli/gravitino problem \cite{Kohri:2005ru}.
Here we considered a more general set-up based on 4D effective
SUGRA with SUSY breaking uplifting potential,
and noted that the scheme can result in a highly distinctive
superparticle spectrum at low energy scales.
This can be easily understood by noting that the low energy soft parameters
in a mixed modulus-anomaly mediation with  messenger scale
$\Lambda$ are (approximately) same as those
of the pure modulus-mediation with
a {\it mirage messenger scale} $\sim (m_{3/2}/M_{Pl})^{\alpha/2}\Lambda$
where $\alpha= m_{3/2}/[M_0\ln(M_{Pl}/m_{3/2})]$
for $M_0$ denoting the modulus-mediated contribution to
the gaugino mass at $M_{GUT}$.
The minimal KKLT model predicts $\alpha=1$, thus has a mirage messenger scale
close to the intermediate scale $\sqrt{m_{3/2}M_{Pl}}$,
while the string, compactification and gauge unification
scales are all close to $M_{Pl}$.
The most dramatic situation is $\alpha=2$ for which
soft masses appear to be unified at TeV
although the gauge couplings are unified at  $10^{16}-10^{17}$ GeV.
Although no string theory realization is found yet,
$\alpha=2$
can be naturally obtained by an uplifting mechanism to yield
an uplifting potential $V_{\rm lift}\propto 1/(T+T^*)$ \cite{choi1,choi2}.
Alternatively, one might be able to obtain such a value of $\alpha$
by tuning the form of the non-perturbative superpotential \cite{choi2}.
All the results of our phenomenological analysis are summarized in
Figs. 1--\,\ref{fig:hierarchy} and Table 1.

\bigskip
{\bf Acknowledgments}

\medskip

K.C. would like to thank A. Falkowski and H. P. Nilles for helpful
discussions. This work is supported by KRF PBRG 2002-070-C00022, the BK21 program of
Ministry of Education and the Center for High Energy Physics of Kyungbook National University.


\listoftables           
\listoffigures          

\end{document}